# Forecasting asylum-related migration flows with machine learning and data at scale

Marcello Carammia[1,2], Stefano Iacus[3], Teddy Wilkin[2]

## Summary

The effects of the so-called 'refugee crisis' of 2015-16 continue to dominate the political agenda in Europe. Migration flows were sudden and unexpected, leaving governments unprepared and exposing significant shortcomings in the field of migration forecasting. Migration is a complex system typified by episodic variation, underpinned by causal factors that are interacting, highly context dependent and short-lived. Correspondingly, migration monitoring relies on scattered data, while approaches to forecasting focus on specific migration flows and often have inconsistent results that are difficult to generalise at the regional or global levels.

Here we show that adaptive machine learning algorithms that integrate official statistics and non-traditional data sources at scale can effectively forecast asylum-related migration flows. We focus on asylum applications lodged in countries of the European Union (EU) by nationals of all countries of origin worldwide; the same approach can be applied in any context provided adequate migration or asylum data are available.

We exploit three tiers of data – geolocated events and internet searches in countries of origin, detections of irregular crossings at the EU border, and asylum recognition rates in countries of destination – to effectively forecast individual asylum-migration flows up to four weeks ahead with high accuracy. Uniquely, our approach a) monitors potential drivers of migration in countries of origin to detect changes early onset; b) models individual country-to-country migration flows separately and on moving time windows; c) estimates the effects of individual drivers, including lagged effects; d) provides forecasts of asylum applications up to four weeks ahead; e) assesses how patterns of drivers shift over time to describe the functioning and change of migration systems.

Our approach bridges the worlds of migration theory and modelling, international protection, and data science to deliver what is, to our knowledge, the first comprehensive system for forecasting asylum applications based on adaptive models and data at scale. Importantly, this approach can be extended to forecast other social processes.

---

[1] University of Catania. Via Vittorio Emanuele II, 49, 95125 Catania, Italy.
[2] European Asylum Support Office (EASO). MTC Block A, Winemakers Wharf. Grand Harbour Valletta, MRS 1917, Malta.
[3] European Commission, Joint Research Centre. Via Enrico Fermi, 2749, 21027 Ispra (VA), Italy.

## Introduction. Problem and background

The 2015-16 refugee crisis in Europe was sudden and unexpected, and the actions taken by Governments to secure borders and uphold access to asylum procedures were generally reactive, uncoordinated and ineffective. The ultimate cause of the crisis was poor capacity to anticipate the movements of asylum seekers. Here we demonstrate that unsupervised adaptive machine learning algorithms can combine administrative and non-traditional data at scale to deliver short term forecasts of flows of asylum seekers from any country of origin to European Union (EU) countries – and in principle to any country that collects data on asylum applications with adequate frequency.

Forecasting asylum-related migration is extremely problematic. Migration is a complex system,[1] which means that causal factors interact nonlinearly, are highly context dependent, and show little or no persistence over time. Potential drivers are diverse,[2,3] plus effect sizes and interactions vary wildly between and within individual migration flows. In one context extreme conflict, violence and persecution may generate few asylum seekers; whereas elsewhere relatively subtle social unrest may spark large international displacements, particularly if they are a tipping point of deteriorating conditions. The effect of migration drivers is subject to threshold and feedback effects. Once activated, country to country flows tend to trigger self-reinforcing processes resulting in the establishment of migration systems [4–6].

Asylum related migration is therefore a highly uncertain process,[7] which complicates migration modelling.[8] Among migration types, forced or asylum-related migration is associated with the highest uncertainty.[9] As a consequence, quantitative asylum migration models often focus on single drivers in countries of origin (e.g. conflicts[10–12]) or destination (e.g. migration or asylum policies[13–15]). More comprehensive asylum migration models have been developed but these aim to increase retrospective understanding rather than forecasting flows[10,16–20], with exceptions mostly confined to the prediction of single country to country flows[21].

Data on migration in general and its drivers also contain uncertainty, which further complicates migration modelling[7]. Despite recent advances in the collection of official statistics, particularly in the subfield of asylum, and in spite of the ongoing efforts to improve data collections at the international[4] and global[5] levels, most data collections are limited in terms of frequency, definitions, coverage, accuracy, timeliness, and quality assurance[22–24]. This is also the case for data on migration drivers such as conflicts, the state of human rights and the economy – notably with regards to their frequency, accuracy and timeliness – all of which are prerequisites for effective forecasting.

Recent developments in data at scale and computational technologies underpin some progress in asylum migration modelling and forecasting. Large data sets containing vast reams of structured and unstructured data have been proposed as an opportunity to observe the unbroken totality of potential migration drivers as they occur in near to real time[25,26]. New data sources include mobile data[27], social media[28,29], and internet searches[30]. Big data are increasingly analysed with models such as agent-based modelling[31] and techniques such as machine learning[32] to detect patterns and identify potential migration drivers that would otherwise go unnoticed. Such developments enabled the development of novel migration forecasting models, including forced and asylum migration, with encouraging results in terms of reliability and timeliness – which makes them potentially useful in operational scenarios[32,33]. However, to our knowledge even the most advanced models have been applied to a limited number of flows rather than generalised to the regional or global levels.

---

[4] Notably in the European Union. See Eurostat's Migration and population statistics at
https://ec.europa.eu/eurostat/statistics-explained/index.php?title=Migration_and_migrant_population_statistics;
European Asylum Support Office Statistics https://easo.europa.eu/analysis-and-statistics; and the Knowledge Portal of the European Commission's Knowledge Centre on Migration and Demography (KCMD) at
https://bluehub.jrc.ec.europa.eu/portal/ (all accessed on 18 August 2020).
[5] See UNHCR Global data at https://migrationdataportal.org/institute/unhcr-global-data-service and IOM Global migration data portal at https://gmdac.iom.int/global-migration-data-portal (all accessed on 18 August 2020).

# Design

## Approach

Here we show that adaptive machine learning algorithms that integrate official statistics with non-traditional data sources can effectively capture early warning signals of asylum-related migration and forecast asylum applications in countries of destinations. Our early warning and forecasting system proceeds by monitoring countries of origin to detect drivers of migration early onset; estimating the effects of individual drivers, including lagged effects; on those bases, producing forecasts of asylum applications up to four weeks ahead; finally, assessing how patterns of drivers shift over time to describe the functioning and change of migration systems.

To observe migration covariates at different stages of migration processes, we exploit three tiers of data: geolocated events and internet searches in countries of origin; detections of irregular crossings at the borders of the EU; and asylum recognition rates in countries of destination. We leverage the potential of non-traditional data sources to position the analysis as close as possible to migration drivers as they happen.

Our approach addresses the complexity of migration systems and drivers by modelling each single country-to-country migration flow separately. Moreover, models are trained on moving time windows to account for change over time even within individual flows. We applied our method on a wide range of bilateral asylum flows covering all (circa 200) countries of origin worldwide, and 30 countries of destination in Europe. In the Extended Data section we report performance results for 70 country-to-country flows, generated by seven countries of origin (Afghanistan, Eritrea, Iraq, Nigeria, Syria, Turkey and Venezuela) and nine countries of destination (Austria, Belgium, Germany, Greece, Spain, France, Italy, The Netherlands and Sweden), plus the EU+ as a whole. The selected flows represent a suitably large diversity on the variables analysed; but our method can be applied in any context, provided adequate migration or asylum data are available.

The workflow of the early warning and forecasting system is sketched in Figure 1.

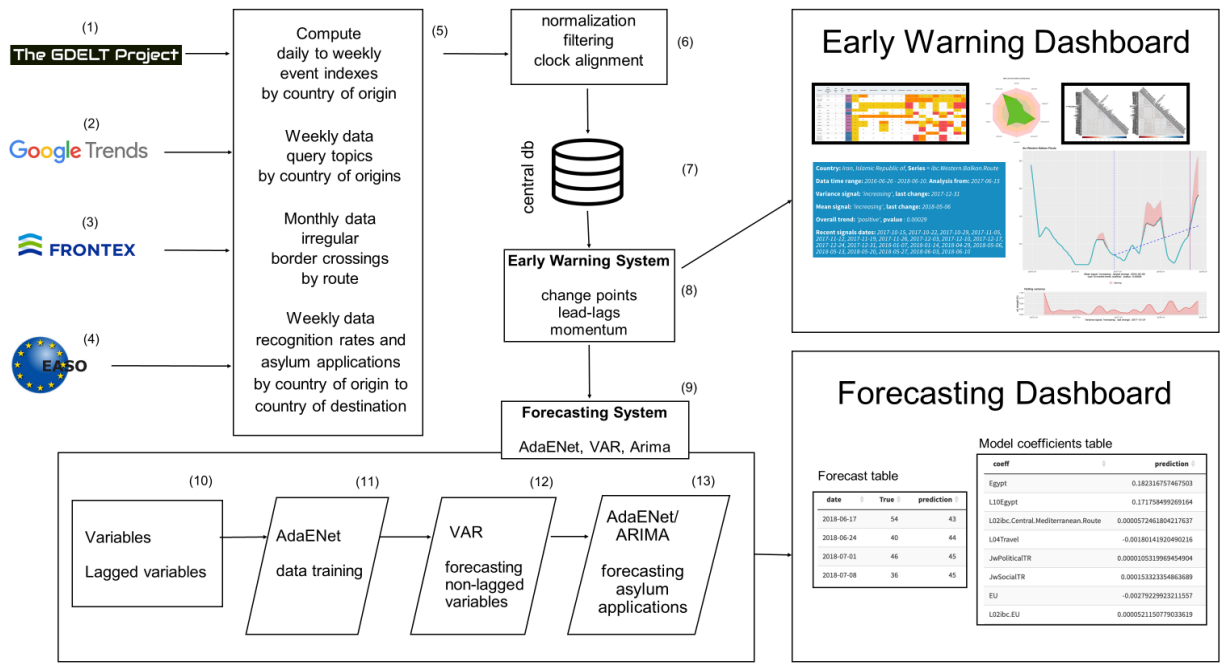

*Figure 1. Overview of the Early Warning and Forecasting System workflow.*

## Target variable

Our dependent variable is the number of asylum applications lodged in each country member of the EU Common European Asylum System (which includes EU Member States plus Norway and Switzerland, and is referred to as EU+), and in the EU+ as a whole [Figure 1-(4)]. These data, broken down by nationality of each applicant, are shared with the European Asylum Support Office (EASO – the EU asylum agency) on a weekly basis. Although provisional, in 2019 these data underestimated asylum applications by just 6% at the EU+ level, compared to official national statistics compiled by Eurostat (EU 862/2007). Estimates of asylum applications typically have a maximum value imposed upon them by the capacity of the receiving country to react to sudden influxes and quickly register each application. This ceiling effect is demonstrated by fewer applications being registered during Christmas and Easter holidays due to the brief closure of some asylum offices, although applications tend to be registered shortly thereafter[34].

## Covariates

To measure covariates we combine administrative statistics with non-traditional data.

### TIER1: Data in countries of origin

### Events

We estimate the timing and location of 'push factor' events in nearly all countries of origin from the Global Database of Events, Language, and Tone (GDELT) project, [Figure 1-(1)], a repository of 316 types of geolocated event reported in the world's broadcast, print and web media, in 100 languages. [6] We used GDELT 1.0 data, which are updated on a daily basis; GDELT 2.0 data are updated every 15 minutes and could potentially feed a real-time system. Single events can be covered multiple times by different media outlets around the world and therefore occur multiple times in the GDELT data, so we extract individual events from the overall media coverage and include each event only once in our data. Not all events reported by the media are expected to be migration drivers and so we selected a subset of 240 events as potential drivers of migration and displacement.[35] Individual events differ in the extent to which they are likely to affect displacement and migration, therefore we weighted (+-) each individual event according to its potential to generate displacement. Then we aggregated all weighted events into five macro-categories: *political events*, *social unrests*, *conflicts*, *economic events*, *governance-related events* [Figure 1-(5)]. Finally, the total number of weighted events per macro-category per week were used in the following models, in order to align them with the frequency of the response variable. More details on event selection and weighing and on the general construction of event indexes are presented in the Methods section. The complete list of events and weights is included in the Supplementary Information section.

---

[6] See Web Page at: https://www.gdeltproject.org. GDELT "monitors the world's broadcast, print, and web news from different countries in over 100 languages and identifies the people, locations, organizations, counts, themes, sources, emotions, counts, quotes, images and events driving society daily".

Internet searches for particular topics may anticipate migration, and indeed internet searches have recently been used to estimate migration intentions and predict migration flows[30]. To estimate patterns of relevant internet searches, we use Google Trends, a publicly available search engine providing multi-language, geolocated data on the relative frequency of search topics and terms[7] [Figure 1-(2)].

As shown in Table 1, we selected 17 topics (clusters of keywords) related to international migration and travelling in general (*visa* or *passport*), asylum seeking (*right of asylum* or *refugee*), countries of transit (e.g. *Jordan* or *Turkey* for searches that take place in Syria) and countries of destination (e.g. *Germany*, *France*, or *EU*). Then we downloaded the relative search frequencies for these topics in non-EU countries. The General User Interface of our system permits to easily customise the topic searches to include in the analysis of single countries. For example, we included searches related to some particular 'transit countries' among search topics (see Table 1). We selected those particular countries of transit for inclusion among topic searches because in the period covered by this analysis the arrivals were high on the Greek-Turkish border, and countries in the Middle East and Central Asia were the origin of a significant share of asylum applicants in Europe. Our model would then discard those variables if not relevant as predictors (for example, while performing forecasts of asylum applications by nationals of Venezuela). In principle, we could have included different sets of topic searches for each individual country of origin included in the analysis.

*Table 1. Google Trend topics (not keywords)*

| MIGRATION/TRAVEL | ASYLUM | TRANSIT | DESTINATION |
|---|---|---|---|
| Passport | Refugee | Egypt | Cyprus |
| Travel | Right of Asylum | Iraq | France |
| Travel visa | | Jordan | Germany |
| | | Lebanon | Greece |
| | | Turkey | Italy |
| | | | Spain |
| | | | European Union |

---

[7] See the Google Trends search engine, available at www.trends.google.com.

*TIER2: Data from the EU external border*

*Detections of irregular border crossing*

EU Member States and Schengen Associated Countries share monthly detections of illegal border-crossing with the European Border and Coast Guard Agency (Frontex) who make the data available on their website, aggregated at the level of 'migration route' (Eastern, Central and Western Mediterranean; Western Balkans, Eastern Borders) [Figure 1-(3)][8]. We are not aware of systematic research on the relationship between illegal border-crossings and asylum applications (but see[13]) but EASO's unpublished analyses suggest that the relationship interacts with nationalities and locations. For example, irregular border crossings do not precede asylum applications for nationalities that can travel to the EU via regular means i.e. with visa or under visa-free regimes, but they do precede asylum applications in locations where detection at the external border is inevitable, such as the Greek Aegean islands. In any case, we include illegal border crossings in all our models because our machine learning algorithm retains variables relevant to the individual flows and discards those that are not.

*TIER 3 – Asylum processes data in the EU+*

To capture the potential effect on asylum applications of asylum processes and practice in countries of destination, we include data on 'recognition rates' [Figure 1-(4)] calculated as the share of total asylum decisions that grant (rather than reject) international protection.[9] EU+ Member States and Schengen Associated Countries share monthly asylum decisions with EASO broken down by the nationality of the applicant; validated data are later released by Eurostat. Recognition rates vary markedly between receiving countries, and they have been shown to be positively related to asylum applications[15]. Moreover, recognition

---

[8] Data are defined by EBGCA as "data reported on a monthly basis by Member States and Schengen Associated Countries on detections of illegal border-crossing on entry between Border Crossing Points of the external borders of the Member States of the EU and Schengen Associated Countries, and aggregated by routes". As EBCGA reports, "The data refer to detections of illegal border-crossing rather than the number of persons, as the same person may cross the external border several times. However, there is currently no EU system in place capable of tracing each person's movements following an illegal border-crossing. Therefore, it is not possible to establish the precise number of persons who have illegally crossed the external border." (source: notes to the data spreadsheet downloaded at https://frontex.europa.eu/along-eu-borders/migratory-map/, last checked on 17 February 2020).

[9] Refugee status or Subsidiary protection status as defined in Article 2 of the Qualification Directive 2011/95/EU

rates can cause a deflection effects[13,16] whereby lower recognition rates in one country may induce asylum seekers to lodge their applications in countries with higher recognition rates.

## Procedure

One of the most severe constraints to migration theory and modelling is that migration processes connect single country of origin to country of destination dyads in complex systems whose functioning vary largely over time and space. To address this constraint, rather than attempting to build a single asylum migration model we model each individual country-to-country dyad separately. In practice, the procedure starts with the selection of one or more countries of origin and proceeds with the analysis of time series on three tiers of potential covariates:

1. In countries of origin:
   - events (5 macro-categories)
   - internet search queries (17 topics)
2. At the external border of the EU:
   - detections of irregular border crossing (across 4 migratory routes)
3. In countries of destination:
   - recognition rates (in 30[10] EU+ countries individually and in the EU+ as a whole)
   - asylum applications in all EU+ countries and in the EU+ as a whole

Given that uncertainty increases with the forecasting horizon, we forecast the number of asylum applications (one, two, three, and) four weeks ahead. While a relatively short term, four weeks is a highly valuable time window for planning and preparedness purposes in an operational context.

We design a system that works in two steps: early warning and forecasting.

---

[10] In the period covered in the analyses, the United Kingdom was still a Member State of the European Union and the Common European Asylum System.

In the first step – **early warning analysis** [**Figure 1-(8)**] – for each country of origin, the system takes all variables and detects signals of significant change by performing change point analysis for mean and variance in each time series. Then Pearson correlations are performed between all covariates and asylum applications lodged by nationals of each country of origin; , and the lags that maximise correlations between each pair of series is estimated.

This system is entirely data-driven. The activation thresholds to trigger alerts depend on a moving average window of the latest data available. Single countries of origin have different "natural" levels of conflicts and other potential migration-generating events, different patterns of internet searches, generate different volumes of asylum applicants, and so forth. Fixed thresholds may result in inconsistent false positive alarms. The optimal predictive lag of each time series on the outcome variable – asylum applications – is found using a lead-lag estimation method[36]. In addition to lead-lag analysis and change point estimation, the acceleration of each time series is measured based on the ratio between shorter (6 weeks) and longer (24 weeks) moving averages. We borrow this 'momentum approach' from quantitative finance[37], and set at ± 110% the ratio's threshold for triggering alerts (cf. [38,39]), but like the other parameters of the early warning system this one also can be customised. The early warning analysis is explained in full details in the Methods section.

Figure 2 shows an illustrative early warning summary ran in the week ending on 17 June 2018 with a focus on the following countries of origin: Afghanistan, Iran, Iraq, Albania, Eritrea, Georgia, Nigeria, Pakistan, Russia, Syria, Turkey, Venezuela (with a closer focus on the particular the case of Iran).

**a**

| Country | Last Month Applicants EU | Tot Alerts EU | Last Alerts EU | Trend EU | pvalue | JwConflict | JwGovernance | JwPolitical | JwSocial | JwEconomy | Germany | Italy | Greece | France | Spain | Passport | Travel | Refugee | EU |
|---|---|---|---|---|---|---|---|---|---|---|---|---|---|---|---|---|---|---|---|
| Afghanistan | 4115 | 6 | 0 | negative | L1 | L2 | L0 | L0 | L0 | L0 | | | | | | | L1 | L1 | L0 |
| Syrian Arab Republic | 6199 | 6 | 0 | negative | L0 | L0 | L0 | L0 | L0 | L0 | L2 | L2 | L2 | L2 | L1 | L1 | L1 | L3 | L0 |
| Iran, Islamic Republic of | 1662 | 21 | 2 | positive | L0 | L1 | L1 | L1 | L0 | L1 | | | | L1 | | | | L3 | |
| Iraq | 3291 | 7 | 7 | negative | L1 | L1 | L0 | L0 | L0 | L0 | | | | | L1 | | | L0 | |
| Turkey | 1939 | 32 | 17 | positive | L1 | L1 | L0 | L1 | L0 | L0 | L1 | | | | L1 | | | L1 | |
| Albania | 1690 | 20 | 1 | negative | L1 | L0 | L2 | L2 | L0 | L2 | | | | | L1 | | L1 | L1 | |
| Eritrea | 1666 | 9 | 3 | negative | L0 | L0 | L0 | L0 | L0 | L0 | | | | L1 | | | | | L0 |
| Nigeria | 2620 | 7 | 0 | negative | L1 | L1 | L0 | L2 | L0 | L0 | L2 | L2 | L1 | L2 | L1 | L2 | L2 | L1 | |
| Pakistan | 2654 | 7 | 0 | negative | L1 | L1 | L0 | L1 | L0 | L0 | | | | L1 | | | L2 | L1 | |
| Russian Federation | 1474 | 13 | 0 | negative | L1 | L1 | L0 | L0 | L0 | L0 | | | | L1 | | | L1 | L1 | |
| Georgia | 1516 | 54 | 28 | positive | L1 | L1 | L0 | L0 | L0 | L0 | | | L0 | L2 | L2 | | L1 | L1 | L0 |
| Venezuela, Bolivarian Republic of | 3455 | 55 | 31 | positive | L1 | L1 | L0 | L0 | L0 | L0 | L0 | L0 | L0 | | | | | | |

**b**

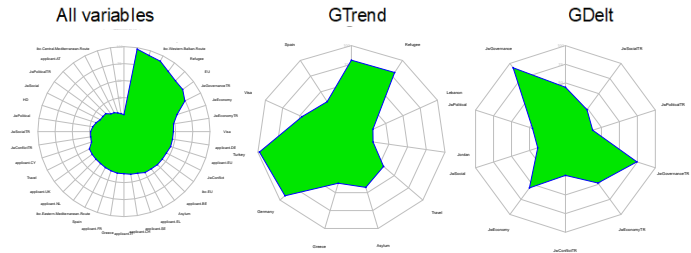

All variables    GTrend    GDelt

**c**

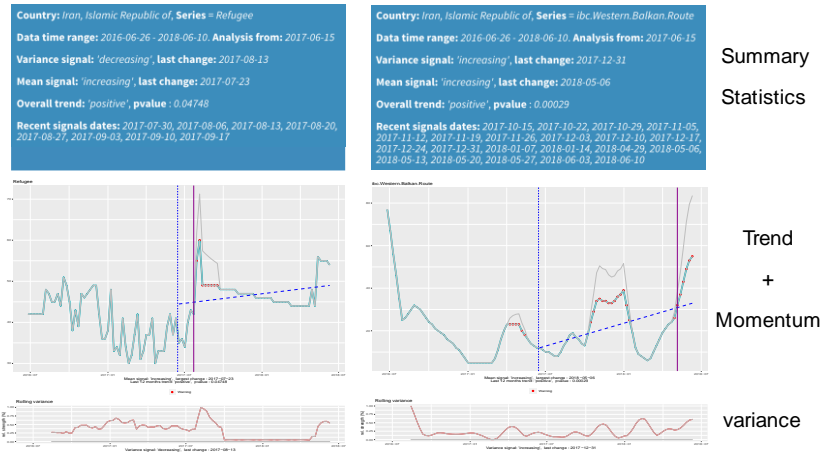

Summary Statistics

Trend + Momentum

variance

**d**

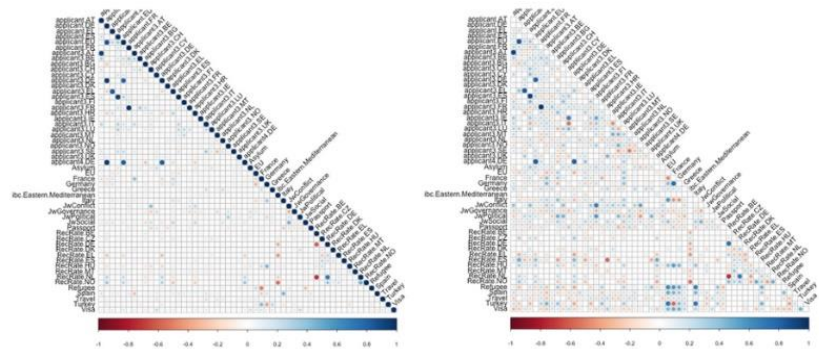

***Figure 2. Early Warning Summary, week ending on 17/06/2018.***
*a. Early Warning Signals Table. Countries of origin included (rows): Afghanistan, Iran, Iraq, Albania, Eritrea, Georgia, Nigeria, Pakistan, Russia, Syria, Turkey, Venezuela. For each country, the table first shows the total number of applicants, the number of signals observed and the trend of applications in the EU+ in the previous month. The degree of warning for each covariate (columns) is then shown: L0 (no warning) to L3 (max warning). Covariates included in the table are event macro-categories (conflicts, governance, political events, social unrest, economic events) and Google Trends topics [searches related to countries of destination (Germany, Italy, Greece, France, Spain, EU) and migration (passports, travel, refugee)]. The table highlights the time series that deserve closer inspection.*

**b**. *Iran, week ending on 15/06/2017. Radar plot of relative level of activity of single covariates in the early warning window, here set as one month, compared to the entire period of analysis: GDelt events and Google Trends searches, level during the early warning window relative to each series' past values (left); Google Trends relative volume of searches (middle). Gdelt event indexes relative level of activity (right). All series rescaled to 0-100%.*
**c**. *Iran, week ending on 15/06/2017. Time series with signals for individual covariates. In this figure: Google Trends searches for "Refugee" topic in Iran and Frontex's "Irregular Border Crossings at the Western Balkan Route" of Iranian nationals. Top of each panel: the summary statistics, recent "momentum" signals, change point statistics for the mean and the variance. In the middle panel: data for the early warning windows with signals and change point analysis. Bottom panel: cumulative rolling variance to check for instability of the time series.*
**d**. *Iran, week ending on 15/06/2017. Correlation matrices, with and without shifting the time series for the optimal lag. At the optimal lag many correlation effects emerge.*

## Forecasting

Also based on the statistics generated in the early warning step, in the **forecasting** step [Figure 1-(9)] the system estimates the future number of asylum applications in European countries of destination, aggregated by the nationality of the applicant. Our machine learning algorithm uses a rolling window of past data on single country-to-country dyads, including lagged covariates identified in the early warning step, to model those processes and then generate projections [Figure 1-(10)].

More specifically, for each country-to-country dyad, the procedure consists in estimating an Adaptive Elastic Net Model (AdaENet, see Methods section) on a moving time window of historical data. The procedure uses the 12 months preceding the observation point as a training set to estimate an AdaENet model, as well as for further cross validation to minimize the Mean Squared Error (MSE) of the forecasts across the training period. Being an adaptive method that mixes LASSO-type and Ridge-type estimation, the AdaENet takes into account hundreds of variables for each dyad of country of origin and country of destination and it has the advantage of finding the most parsimonious model for each dyad, i.e., it performs model selection and estimation contextually. At the same time, the model also takes into account collinear variables as in Ridge regression [Figure 1-(11)].

A mixed strategy of VAR and ARIMA modelling is used to predict the future values of the covariates (see Methods section for details). For lagged variables, the real past values are considered as predictors [Figure 1-(12)]. Finally, when all predictors have been forecasted, future values of the applications lodged at 1, 2, 3 and 4 weeks are obtained by feeding the forecasted predictors into the AdaENet selected model. The optimal (see Methods section) AdaENet model is then bench-marked against the Arima(1,0,1) model based solely on asylum applications.

The procedure for the forecasting step is explained in detail in the Methods section, and most parameters can easily be customised. By way of illustration, Figure 3 shows a sample forecast for asylum applications lodged by nationals of Afghanistan in the EU+ in early 2019. The real number of asylum applications is represented by the green line before the point when the forecast is simulated, and by the dotted blue line after that point. The four-week forecast is the red line.

Afghan nationals had been among the top-three nationalities for asylum applications in the EU+ for most of the time between 2014 and 2018. As was the case with most top-ranking nationalities, the volume of applications by Afghani nationals fell sharply after the EU and Turkey agreed to end irregular migration from Turkey – the main country of transit at that time – to the EU in March 2016.[11] In addition to this structural change, the flow was also subject to cyclical movements and some seasonality – notably a drop at the end of each year when the processing capacity of asylum authorities tends to be severely limited, followed by subsequent increase. Both AdaENet and Arima attempt to capture this behaviour, but AdaENet is far more effective. In the four weeks considered in the back-test, the total number of applications was 3424; AdaENet forecasted 3445 (0.6% relative error, 21 units absolute error) and Arima 2826 (-17.5% relative error, 598 units).

---

[11] See https://ec.europa.eu/commission/presscorner/detail/en/MEMO_16_963. As a result, Irregular crossings at the EU border by Afghan nationals went down from a monthly average of about 26700 the year before, to one of about 1430 the year after. Asylum applications decreased markedly as a result, although the change in the asylum trend became visible some six months after the EU-Turkey statement. Between October 2015 and September 2016, the average number of monthly asylum applications lodged in the EU+ was 123308, which went down to a monthly average of 63170 between October 2016 and September 107.

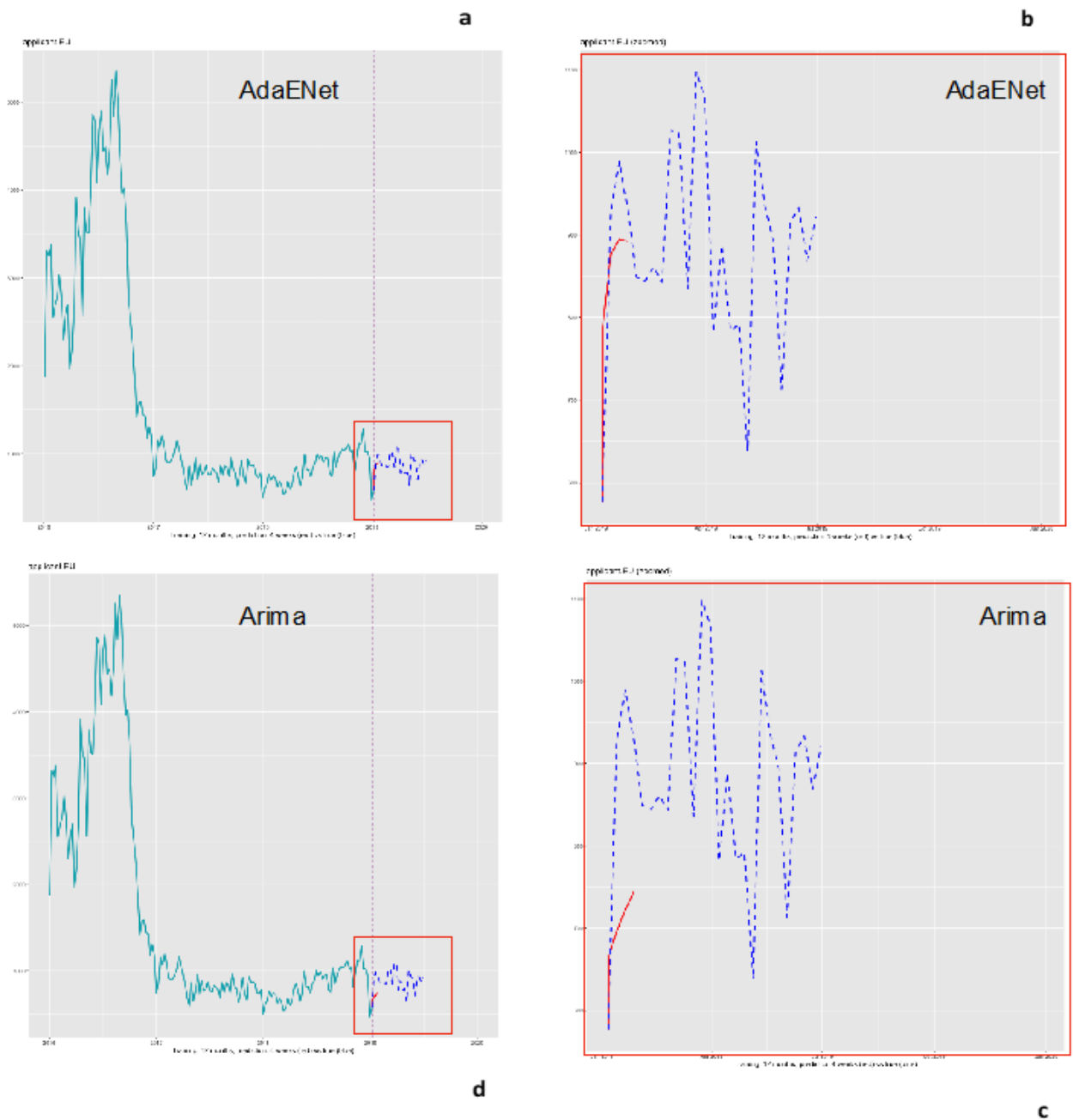

*Figure 3. Forecast of applications by Afghan nationals in all EU+ for the four weeks following 30/12/2018.*
*a-b: AdaENet model; c-d: Arima model. Figures 3a and 3c show the full series, while figures 3c and 3d zoom in on the period starting with the forecast. Weeks are represented in the x axes, and the number of applications lodged in EU+ countries in the y axes. The green line shows the number of applications lodged until the point in which the forecast is launched. The forecast is represented by the red line. The blue dotted line show the actual number of applications lodged over the forecast period (and afterwards). The chosen week is a very difficult test for both models, as the process has a huge drop down in coincidence with the end of the year when few applications were processed, and rebounds shortly after. The AdaENet model manages to cope better with the anomaly than what Arima can do.*

## Forecasting performance

### *Back-testing and forecasting errors*

To test the forecasting performance, we simulated weekly forecasts from 30 April 2017 to 1 September 2019. The forecasting system takes data from April 2016 to April 2017 and then iteratively moves onward by one week at every step. This means that the procedure replicates a hypothetical real forecast using only information that would have been available at each point in time, each time running early warning analyses to generate lagged variables that can be retained by the system in the forecasting step.

Figure 4 shows the back-testing results for an especially relevant flow: Syrian applicants (SY) in Germany (DE), effectively the largest flow in the EU+ for most of the time between 2014 and 2019. The series shows some typical patterns of asylum processes, such as non-regular cyclical oscillations, as well as some stylized properties of these administrative data such as the drop at the end of each year that has been observed above for Afghans applicants.

In almost all weeks, the forecast stays within the confidence bands of +-2 standard errors, which means that the system performs statistically well. The exceptions are limited and occur mostly during the initial part of the analysis. That is largely due to the structural change in the series following the EU-Turkey statement in March 2016, which resulted in a radical change in the trend of asylum applications (see footnote 10 and Supplementary Information for more details). Asylum-related migration can be a rather unstable process in general; including such a radical change within the training period makes the test extremely challenging. Our model typically adapts to change with a short delay; sometimes, for example around June 2018 and 2019, it manages to capture and anticipate abrupt changes. For the particular case of Syrians lodging applications in Germany, the average and median relative errors are 7% and 4% respectively (4.1% and 2.4% respectively from the moving average). For the benchmark ARIMA model, the average and median relative errors are 15.2% and 14.7%. Our model significantly outperforms the ARIMA model most of the time in most country-to-country flows (see Supplementary Information for more details), which shows the added value compared to time series extrapolation methods based on autoregressive models.

Extended Data Figures 1-10 show the same figures for a selected sample of 10 additional flows.[12]

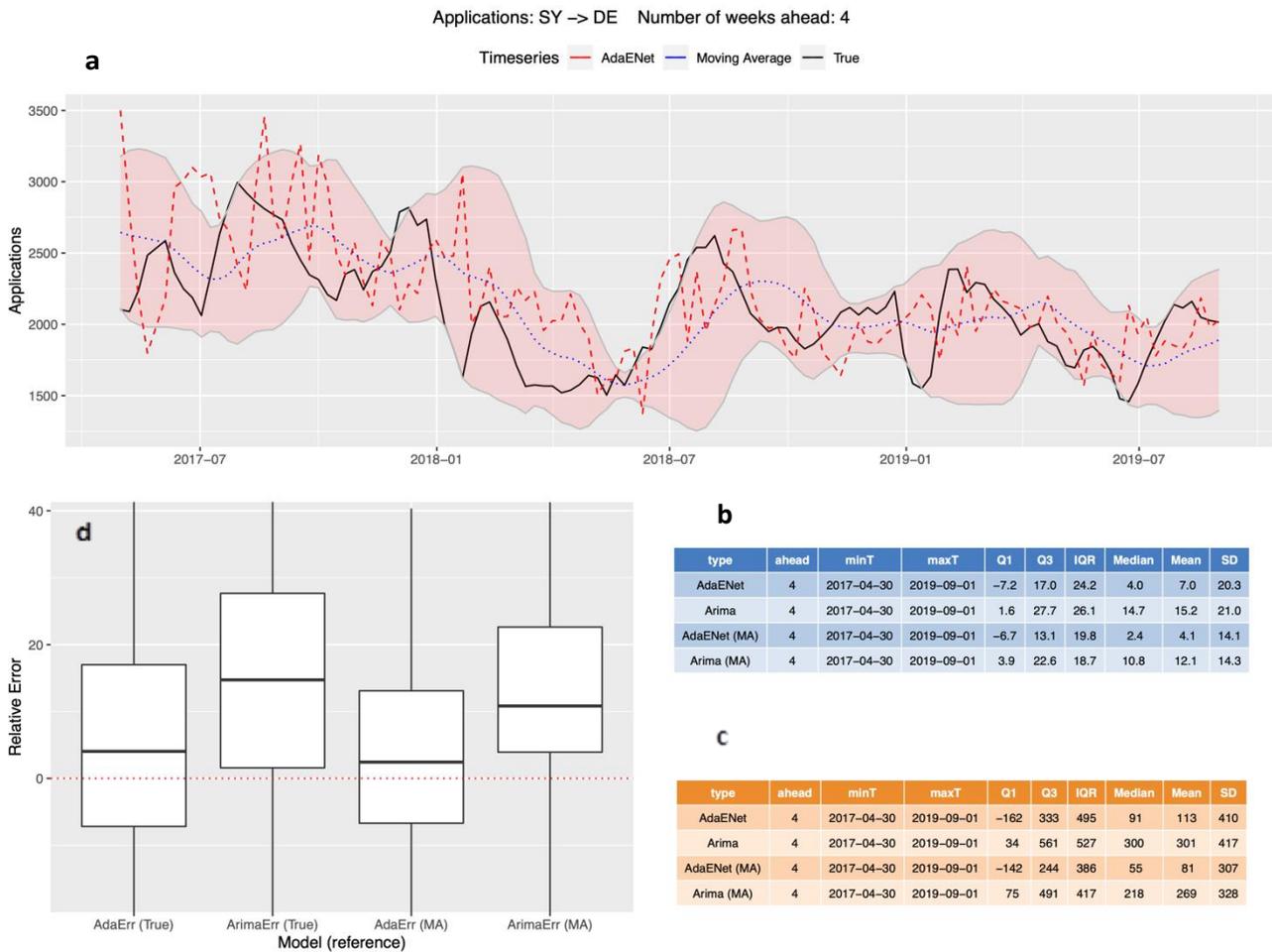

To further illustrate the performance of the system over time and space, in the Supplementary Information section we present an analysis of forecasting performances for a selection of 70 dyads comprising of seven countries of origin (Afghanistan, Eritrea, Iraq, Nigeria, Syria, Turkey and Venezuela)[13] and ten destinations (Austria, Belgium, Germany, Greece, Spain, France, Italy, The Netherlands and Sweden, and the EU+ as a whole). As discussed further in the Supplementary Information section, these dyads were selected to provide a large variation across relevant variables and thus test the performance of the forecasting system over largely different conditions. More extensive analyses have been carried out. [14] However, the Early Warning and Forecasting System analyses all (circa 200) countries of origin and 30 EU+ countries of destination, and the same performance statistics can be generated for all country-of-origin-to-country-of-destination dyads.

---

[12] Applications lodged by Syrian nationals in Greece, Sweden and the EU+; of Venezuelans in France, Spain and the EU+; of Nigerians in Germany and Italy; of Afghans in Germany.
[13] Between 2016 and 2019, applications by nationals of the seven selected countries of origin lodged 1 654 040 applications in the EU+, 47.6% out of a total of 3 473 050 applications received.

*Adaptiveness of the AdaENet model and assessment of change in migration systems*

The matrix in Figure 5 shows the variables selected by the AdaENet model in each week of the back-testing period. The colours represent the relative importance of the predictor in the AdaENet model. Because the model is adaptive, the variables selected as well as their relative importance may vary from week to week. The relative rank is evaluated through a Random Forest algorithm on the restricted model selected by AdaENet and represented through colours in Figure 5 – from 0 ("variable not included") to 1 ("most important").

The heatmap illustrates some key features of the model and of the underlying processes. First, the effects of single variables tend to persist for several weeks (horizontal colour bands). This indicates that country-to-country asylum flows have some (temporal) structure, which the model is able to capture. Variables are clustered in the heatmap in such a way to show this persistence. In Figure 5, variables that were relevant in the first half of the observed time period are clustered in the lower left area, while variables relevant in the second part of the period are clustered in the upper right area.

The variables retained in the model describe the changing nature of the process, and therefore can be used to interpret it. For example, in the first part of the period observed, Syrian asylum applications in Germany

---

were best predicted by such variables as the recognition rate in Greece, searches of Egypt- or refugee-related topics in Syria, or governance events in Syria. The process changed quite markedly in the summer of 2018, when the main predictors became internet searches of Greece, Germany, Iraq, Lebanon, passport- and travel-related topics, as well as lagged searches of Lebanon and refugee-related topics in Syria; economic events in Syria; social, economic or governance events in Turkey, a key transit country; and recognition rates in some other EU+ countries. In general, the variables selected by the model are largely consistent with the nature of single country-to-country flows, and their change over time.

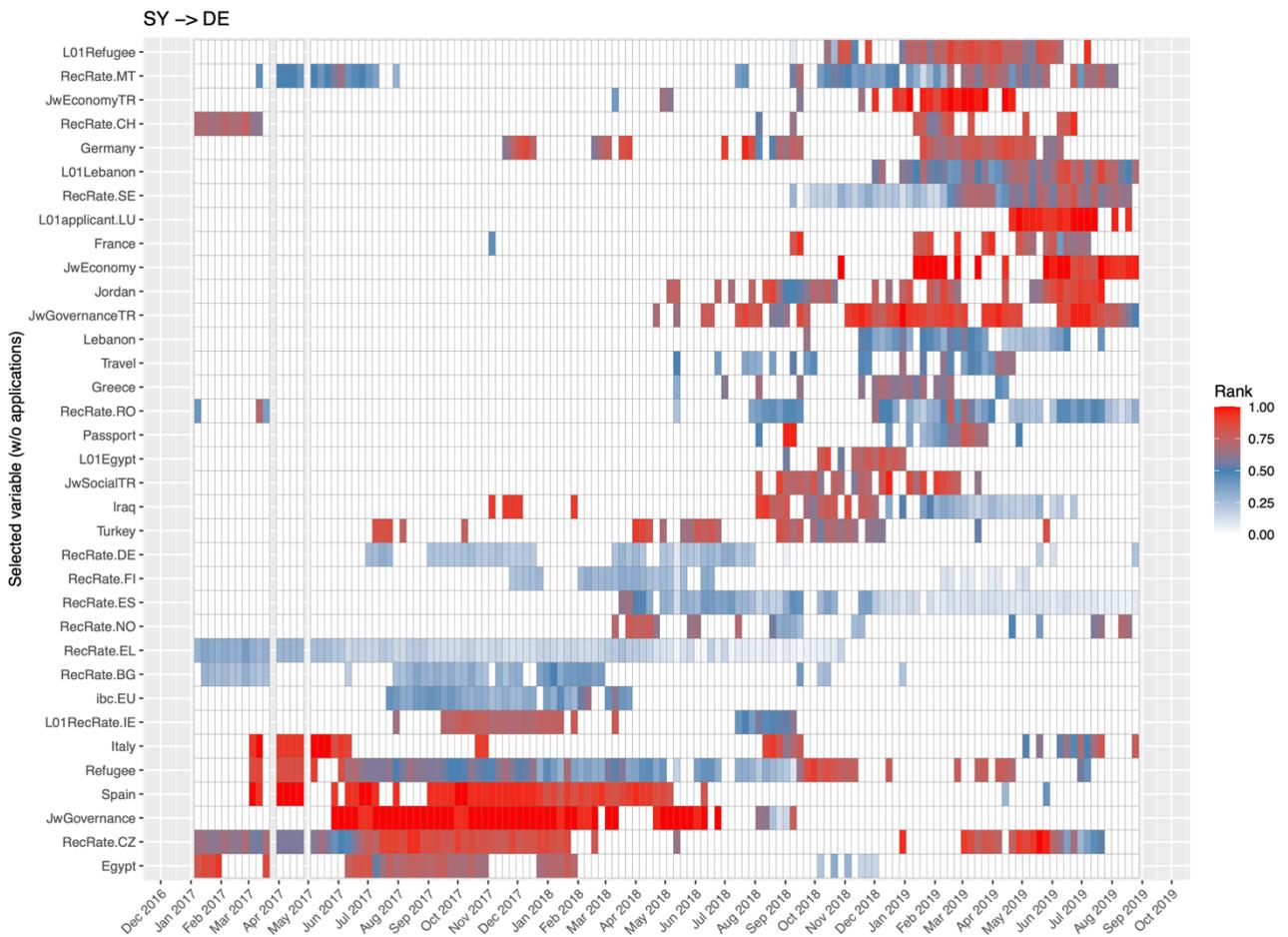

*Figure 5. The factors that impact the forecasting model for Syria towards Germany in the period considered. The model adapts over time, some effects are persistent in the first period, then other become more important. The scale colour only represents the relative importance of the variable. The coloured variables are all included in the model, the others have been dropped by AdaENet.*

Extended Data Figures 1-10 show the same figures for the selected sample of 10 additional flows.

## Conclusion

With increased numbers of displaced persons around the world, irregular migration and asylum have risen up the political agenda. If governments are to effectively manage mixed migration flows, they need to understand cause and effect and plan for future influxes. However, context dependent, short lived complexity combined with sparse data means that forecasts are rare, tentative and unreliable.

This work takes a novel approach to align data on events and internet searches in countries of origin, detections of irregular crossing at the European border, and asylum decisions in European countries. These data are fed into an adaptive machine learning system that delivers short-term forecasts of individual mixed migration flows that consistently outperform benchmark models, as well as assessments of causal factors in the medium to long term. Forecasts of individual migration flows have the potential to inform short term preparedness and resource allocation in the field or in national asylum offices. Assessments of medium and long term drivers contribute to the understanding and modelling of mixed migration flows, and can also be used to inform policy decisions.

Our approach bridges the worlds of migration theory and modelling, international protection, and data science to deliver an effective system for early warning and forecasting of migration. But this approach has the potential to be adapted to understand and forecast other systems and processes characterised by complexity and sparse data.

## Main references

## Methods

### Events indices

We extract from the GDELT database those events that occur in the lead paragraph of a document (coded as 1 on the "IsRootEvent" variable in GDELT). The GDELT project categorises events based on the CAMEO codebook,[15] which includes 316 event categories. Among those, we selected 240 categories that, according to the literature on migration 'push factors',[35][16] are most likely to represent potential drivers of migration. Because the potential to act as a driver varies across single events, we assigned a weight to each of the 240 different event topics, of the type $w = pt/3$, where pt is a number between -4 and 4. The weight $w$ is positive when the event is considered to potentially induce migration; $w$ is negative when the event is considered to potentially reduce or constrain migration. We then aggregated single weighted events in five macro-categories: political events, social unrests, conflicts, economic events, governance-related events. The weights of each event within each macro-category are finally summed to obtain one index per macro-category. Having counted the number of unique (weighted) events for each day, we then aggregate them by week to align them with the frequency of the response variable (asylum applications): `JwSocial`, `JwEconomics`, `JwPolitical`, `JwConflict`, `JwGovernance`.

The CAMEO codebook includes some variables to account for the severity of events. One is QuadClass, which is used to classify events on a 1-4 scale (verbal cooperation, material cooperation, verbal conflict, material conflict). Another one is the Goldstein scale[40], which assigns a numeric score ranging from -10 to +10 based on the theoretical potential impact of single events on country stability. Aggregated in a single composite indicator, that we call Push Factor Index (PFI), our indicator is strongly correlated to the negative Goldstein scale (.9). We tested the correlation between yearly values of the PFI in 2016, 2017 and 2018, on the one hand, and recognition rates as well as one-year lagged asylum applications, on the other hand. On the average, Pearson correlation was respectively 0.54 and 0.47. Taking all events with a negative Goldstein scale, the average correlation with recognition rates and lagged asylum applications was slightly lower – respectively 0.51 and 0.42.

Because it is grounded on migration theory and should therefore reduce the level of noise in the indicators, and because it also seems more closely correlated to asylum applications and recognition rates, we prefer our event indexes. The complete list of GDELT events and those selected for our indicators are available in the Supplementary Information section.

## Internet search queries

We rely on the Google Trends service to crawl the relative search of each topic (see Table 1) for all the countries in the world on a weekly basis.

## Forecasting procedure. Early warning function

### Statistical data cleaning and filtering

A preliminary analysis of each time series is performed in order to drop from the analysis any time series that has insufficient variability (i.e. no statistical information), which can be due either to too low values or to too many missing values as can occasionally be the case for applications and detections of irregular border-crossing. This stage of the pre-analysis has several filtering parameters as shown in Table M1. Before performing this task, the data are aligned to have a common weekly frequency. GDELT data have daily frequency and can be aggregated at the weekly level. Google Trends and asylum applications data are weekly. Data on irregular border crossings and recognition rates have a monthly frequency and are transformed into weekly series through linear interpolation.

The variability is taken into account in terms of the coefficient of variation (cv = standard deviation/|mean|). Being a pure number, i.e. without measure unit, it allows for variable independent thresholding: a time series whose variability is below 5% (the default value) is assumed to be statistically unreliable.

Time series of asylum applications and border crossings are also dropped if their maximal value in the month preceding the forecast is below the thresholds. We have different thresholds for applications, decisions, pending cases and IBC, as illustrated in table M1.

| Parameter | Default value | Details |
| --- | --- | --- |
| country | No default value | Two digits ISO code for country of origin |
| cv.thr | 0.05 | Threshold on the coefficient of variation. Time series with coefficient of variation below the threshold are excluded from the analysis for this country. |
| ibc.thr | 100 | IBC data threshold: if the maximal value of a specific IBC time series is below the threshold, the related data are dropped from the analysis. |
| applicant.thr | 100 | EPS applicant data threshold: if the maximal value of a specific EPS applicant time series is below the threshold, the related data are dropped from the analysis. |
| pending.thr | 100 | Same as above for the pending time series. |
| decision.thr | 100 | Same as above for the decision time series. |
| na.th | 0.3 | If any time series contains more than na.th*100 missing data, the time series is not reliable enough and hence dropped from the analysis. |
| write.db | FALSE | Should write the result to a data base or on files? Currently, only FALSE is available, apart from a subset of data needed for forecasting which are stored anyway on the backend data base. |
| refDate | Sys.Date() | The final date of the analysis. |
| ma1 | 6 | Length of the first moving average (in weeks) |
| ma2 | 24 | Length of the second moving average (in weeks) |
| ma.th | 1.1 | Threshold of first and second moving average. If ma1/ma2 > ma.th, the signal is fired. |
| clean.w | 6 | Data cleaning threshold, in months. All the dropping/cleaning pre-analysis is done only for the last window of data, i.e. the last 6 months. For example, if the maximal value of the IBC data in the last clean.w months is less than ibc.thr, the time series will be dropped. |
| alert.w | 12 | Reference window to analyse the signals (in months) |
| back.w | 24 | Number of past months to consider in the analysis. |
| pvalue | 0.05 | p-value threshold for assessing statistically significant structural change points in time series |
| llag.th | 0.05 | p-value threshold for assessing statistically significant lead-lag effects |

*Table M1. Parameters for the Early Warning function.*

*Change point analysis*

At this point the algorithm performs two change point tests: one for the mean of the time series and one for the volatility in the last `12` months. Statistical hypotheses testing on the change point are calculated with respect to the *p*-value parameter.

*Spotting acceleration*

Some characteristics of the time series could potentially complicate or bias the early warning analysis. For example, a time series may have a slope (positive or negative) without any statistically significant change point, or a change point may have occurred far in the past. The time series may also be highly volatile and change points detected only due to isolated spikes, which however would not imply any persistent change. To take into account the volatility or the speed of change, we compare a *short* and *long* term time series at MA1=6 months and MA2=24 months period. In a period of stationarity, those time series converge, but in case of positive (negative) acceleration, they diverge. If the MA1/MA2 is larger than a given threshold of 1.1, then a signal is fired by the early warning system. There might be isolated signals (the case of huge but isolated spikes) or a series of consecutive signals. In the latter case, the acceleration in activity/trend can be considered as a real signal. This technique is based on high frequency quantitative finance in fast portfolio optimization called the *momentum approach* [37–39].

*Correlation and lead-lag analysis*

Instead of analysing simultaneous correlations, Google Trends and GDELT data were analysed with a Lead-Lag approach, with one time series thereby anticipating another. The lead-lag effect is commonly applied in financial econometrics. In time series, this notion can be compared to Granger causality (see[41]; for empirical evidence, cf. [42–45]). However, Granger-like approaches face several constraints: 1) the time series must be of the same frequency; 2) the time series must be linear, and 3) testing for causality often leads to bidirectional effects. An additional problem is the Epps effect which states that, as the sampling frequency of time series increases, the empirical correlation is reduced[46]. The Lead-Lag approach overcomes the Epps effect by using Hoffman's Lead-Lag estimator[36] based on the Hayashi-Yoshida asynchronous covariance estimator[47–49]. This permits to apply a Lead-Lag approach to asynchronous, non-linear time series with different frequencies and missing data.

More precisely, let $\theta \in (-\delta, \delta)$ be the time lag between the two nonlinear time series $X$ and $Y$. The approach consists in constructing a contrast function $U_n(\theta)$ which evaluates the Hayashi-Yoshida estimator for the times series $X_t$ and $Y_{t+\theta}$ and then to maximize it as a function of $\theta$. The lead-lag estimator $\hat{\theta}_n$ of $\theta$ is defined as

$$\hat{\theta}_n = arg \max_{-\delta < \theta < +\delta} |U_n(\theta)|.$$

When the value of $\hat{\theta}_n$ is positive it means that $X_t$ and $Y_{t+\hat{\theta}_n}$ (or $X_{t-\hat{\theta}_n}$ and $Y_t$) are strongly correlated, therefore we say "$X$ leads $Y$ by an amount of time $\hat{\theta}_n$", so $X$ is the *leader* and $Y$ is the *lagger*. Vice versa for negative $\hat{\theta}_n$. The assessment of the identified lag is done through a statistical test.

To retain the most significant lagged correlation effects, we also calculate the Lead-Lag-Ratio (LLR). LLR is used when there are two lead-lag effects, one for a positive lag and one for a negative lag, both statistically significant. In this case, the strongest among the two shortest (i.e. close to 0) lead-lag effects is returned by the LLR test statistics. Lead-lag and LLR as well as asynchronous correlation are available at present only through the `yuima` R package[50,51].

## Forecasting procedure. Forecasting function

The forecasting system attempts to forecast the value of the variable *applicants* from a country of origin (CoO) to different countries of destination (CoD) including the EU+ in aggregate. To this aim, the information from the early warning step is also used. This means that this task includes the dropping of some of the variables as explained the above.

The analysis focuses on the variables named applicant.*, where * stands for one of CoD.

When data for "future" applications are available, as in back-testing, the system compares the forecast with the actual data. The complete list of arguments of the function forecast are illustrated in Table M2:

| Argument | Default value | Description |
|---|---|---|
| country | No default value. | ISO 2 digit CoO country code |
| final.date | No default value. | Should be in the format "YYYY-MM-DD" |
| start.date | "2017-01-01" | From where to start the back testing meta-analysis |
| n.ahead | 4 | Number of ahead periods prediction (in weeks) |
| prediction.win | 12 | Data used for the predictive model (in weeks) |
| alpha | 0.5 | ElasticNet parameter (see below) |
| burn | 12 | Number of data used in the local predictive statistical models (in weeks) |

*Table M.2. Parameters for the forecast function.*

The forecasting model is the result of a meta-analysis based on all variables with lead-lag effects that entered and survived the previously exposed early warning step.

The forecasting strategy follows these steps:

- Set up an adaptive elastic net model
- Perform model estimation on a moving window of data
- Select the best (in terms of Mean Squared Error) adaptive elastic net model
- Forecast the covariates for the future periods exploiting also lagged variables
- Apply the estimated model to predict the outcome variable.

*Adaptive Elastic Net Model*

Adaptive Elastic Net (AdaENet) is a relatively new type of regularization method which tries to perform model estimation and model selection in just one run.

Suppose we want to estimate a linear model of the form $y = X\beta + \varepsilon$, where $X$ is the matrix of regressors which includes all the variables from the EWS analysis and $y$ is the dependent variable of interest (i.e., applications in this case). In this application, we have a huge number of regressors and relatively too few observations, which prevent us from estimating a new model. Regularization methods, like AdaENet, are also meant for dimensionality reduction, i.e., they estimate some of the coefficients beta as zero.

$$\min_{\beta_o, \beta} \left\{ \frac{1}{N} \sum_{i=1}^{N} w_i l(y_i, \beta_0 + \beta^T x_i) + \frac{\lambda}{2} [(1 - \alpha)||\beta||_2^2 + 2\alpha ||\beta||_1] \right\}$$

where, $w_i$ are weights for observations ($w_i$=1 by default), $l(\cdot)$ is a loss function, normally the classical least squares contrast function, $\lambda$ is a penalty factor and $\alpha$ is a tuning parameter. For $\lambda = 0$, the formula becomes the usual LSE (least squares estimation) approach. For $\lambda = 1$, this method becomes the so called LASSO regression model, i.e., when trying to minimize the squared residuals from the model, the L1-penalty ( $||\beta||_1$ = sum of the absolute values of the regression coefficients) is added forcing some of the coefficients to be estimated as zero. For $\alpha = 0$, this model becomes the Ridge regression model, i.e. the classical regression with shrinkage for robust error estimation. For $\alpha = 0.5$ the model is simply called Adaptive Elastic Net (AdaENet). Using both L1 and L2 penalty at the same time is a good compromise in terms of prediction. In fact, LASSO regression tends to keep only one among highly correlated subsets of regressors discarding all the others. With $\alpha = 0.5$, the result of the regularization takes into account also the correlation among the regressors and it results in a sort of "mean" effect of all variables that matter even though correlated among them. Notice that AdaENet is also a variance shrinking method, which implies that the standard errors of the coefficients of the selected variables are relatively small compared to, e.g., linear regression.

The AdaENet model itself has two tuning parameters: $\alpha$ and $\lambda$. The first one, $\alpha$, is set to 0.5 in our approach, which means Lasso (L1-penalty) and Ridge (L2-penalty) estimations are equally weighted in the loss function of the optimization problem. The parameter $\lambda$, the adaptive scaling factor for the penalties, is first estimated using cross-validation over a very large set of possible values in order to minimise the forecast MSE in the training data. This 'optimal' forecasting value is then used in the AdaENet penalty function. This procedure is performed every week, so the optimal $\lambda$ changes from week to week [Figure 1-(13)].

The tuning parameter $\lambda$ is quite important. The larger this number, the stronger AdaENet will shrink the estimated coefficients to zero, in a potentially artificial way. To take into account this potential source of bias, the elastic net considers a grid of different values for $\lambda$ and estimates the penalized regression model for a particular choice of $\alpha$ (in our case 0.5). Then $\lambda$ is then automatically selected ex-post by cross-validation.

In practice, what happens inside our forecast function is more complicated than the above classical AdaENet algorithm. In particular, historical data on a time varying window are used to estimate the best AdaENet model, i.e. the one with the cross-validated $\lambda$. Then the window is moved one week ahead, and the estimation procedure is calculated. This analysis is iterated until the current data available inside the forecast function. The final model, i.e. the final $\lambda$, is chosen among those models such that they attain the lowest prediction error (as measured by the mean squared error, MSE): for all models such that the estimated MSE is less or equal the variability of the time series, the optimal $\lambda$ is extracted and then, the lowest $\lambda$ among all lambda's, is used in the subsequent analysis.

In summary, instead of considering the best model for the whole period of data, the forecast function selects the best model for each sub-period (time varying window) and then defines the optimal $\lambda$ as the average and the minimal lambda in all the estimated models. This guarantees that the final forecasting model produces on average and in a robust way the best forecast in all periods. The optimal lambda shrinks the coefficients more than the minimal lambda.

### VAR modelling

To obtain the forecast, the model also needs to simulate the future values of the covariates.

To this aim, a Vector Autoregression (VAR) model is first fitted on the historical values of the past 12 weeks of the covariates retained by the AdaENet model; and then used to forecast future values of those covariates. When VAR model estimates do not converge, individual Arima(1,1) models for each predictor are ran. If the Arima(1,1) estimates fail to converge as well, the average value of the time series is considered. This multiple step approach is necessary because some of the time series may not be well approximated by stationary time series. For lagged variables, the real past values are considered as predictors [Figure 1-(12)].

*Sources of uncertainty*

In practice there are two sources of approximation: the first is the simulation of the future values through a VAR/Arima/AR model and the second one is the prediction error of the estimated model on the historical data. Nevertheless, the forecasting models seem to be able to do a statistically sound job, in the sense that forecasts are generally within the 2 standard errors bands of the moving average process based on historical data (as seen, e.g., in Figure 4).

*Random forest for variable ranking*

To have an additional insight on the relative importance of each variable selected by the AdaENet model, we run ex-post a random forest model and we rank the predictors according to the importance measure of this algorithm. As the number of predictors may vary from week to week, we consider the relative rank. This relative rank generates the colour scale in Figure 5.

## Code

R 4.0.0 language was used for the data cleaning, data mashup, statistical analysis and data visualization. Several R packages have been used. in particular glmnet for the implementation of the Adaptive Elastic Net; yuima for the lead-lag analysis; sde, for the change point analysis as well as the package ts and VAR respectively for the ARIMA and VAR models estimation; randomForest was used to extract the relative importance of each predictor in the model selected by the Adaptive Elastic Net algorithm. The rest of the analysis was produced via own written R code.

## Data availability

The data that support the findings of this study are available from the authors. Restrictions apply to the availability of some of the data, and notably to weekly data on asylum applications. Any request of weekly asylum data should be agreed with the national asylum authorities of EU+ Member State, owner of the data. Weekly asylum data are not validated and are exchanged for analytical purposes by the Member States of the EU Common European Asylum System under the Early Warning and Preparedness System of the European Asylum Support Office (for details, see https://easo.europa.eu/analysis-and-statistics). Validated data are made available by Eurostat normally two months after the reporting period, on the monthly level (see https://ec.europa.eu/eurostat/statistics-explained/index.php/Asylum_statistics). Weekly asylum data are therefore not publicly available and were used under license for the current study.

## Methods references

## Author Contributions

MC and TW conceived the project. MC, SI and TW designed the study. SI engineered the code. MC and SI analysed the data. MC, SI, and TW interpreted the results. MC, SI and TW wrote and revised the article.

SI designed the study and engineered the code when at University of Milan; analysed the data, interpreted the results, wrote and revised the article when at European Commission, Joint Research Centre.

## Competing interest declaration

MC was employed at the European Asylum Support Office (EASO) during the conception, design, and piloting of this project. SI was employed at the University of Milan but has been a consultant at EASO for the design and piloting of this project; he is currently on leave from the University of Milan and employed at the Joint Research Centre of the European Commission. TW is employed at the European Asylum Support Office.

## Correspondence

Correspondence and requests for materials should be addressed to Marcello Carammia at

marcello.carammia@unict.it.

# Extended data

This Extended Data section includes ten figures providing statistics and details about the forecasting performance of the system on selected country-to-country asylum flows. The sample flows were selected to provide a diverse set of contexts covering some major destination countries (France, Germany, Greece, Italy, Spain, and the EU+ in the aggregate) ; some major countries of origin, from different global regions (Middle East, Asia, Africa, South America); relatively more (e.g. Nigeria to Italy) and less (e.g. Syria to Greece) volatile flows. Moreover, the associated forecasts range between very good (Extended Data Figures 3, 4, 6 and 7), good (Extended Data Figures 2 and 5), poor (Extended Data Figure 1), and very poor (Extended Data Figures 8 and 9).

More extensive analyses are summarised in the Supplementary Information section (below).

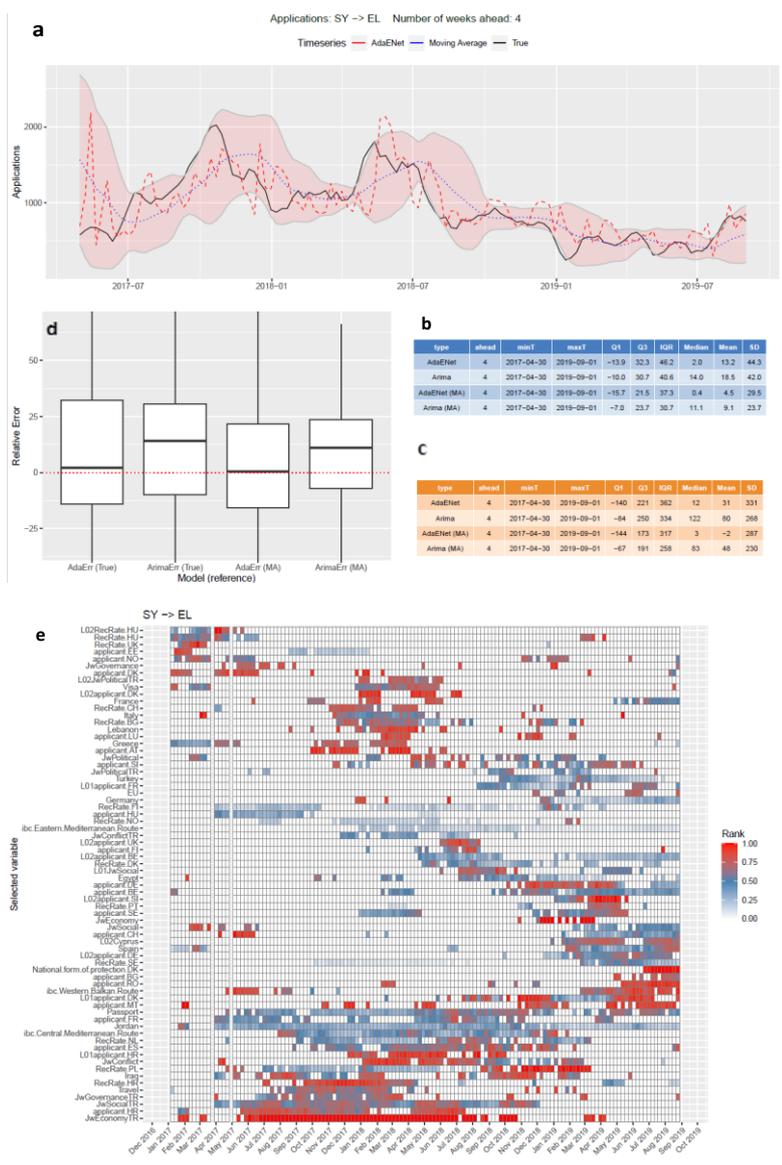

*Extended Data Fig. 1. Back-testing performance of the system for forecasted applications by Syrians in Greece.*
*a. The black line shows the actual number of applications lodged by Syrian nationals in Greece. The dotted blue line is the moving average of the process. The red dashed line shows the AdaENet 4-week ahead forecast at each time point. The pink shaded area represents a ± 2-standard errors confidence bands around the moving average.*
*b-c. Summary statistics for the relative error (b) and for the absolute error (c). Arima is a benchmark model which is only based on the auto-correlation of the application timeseries.*
*e. The factors that impact the forecasting model in the period considered. The model adapts over time, some effects are persistent in the first period, then others become more important. The scale colour only represents the relative importance of the variable. The coloured variables are all included in the model, the others have been dropped by AdaENet.*

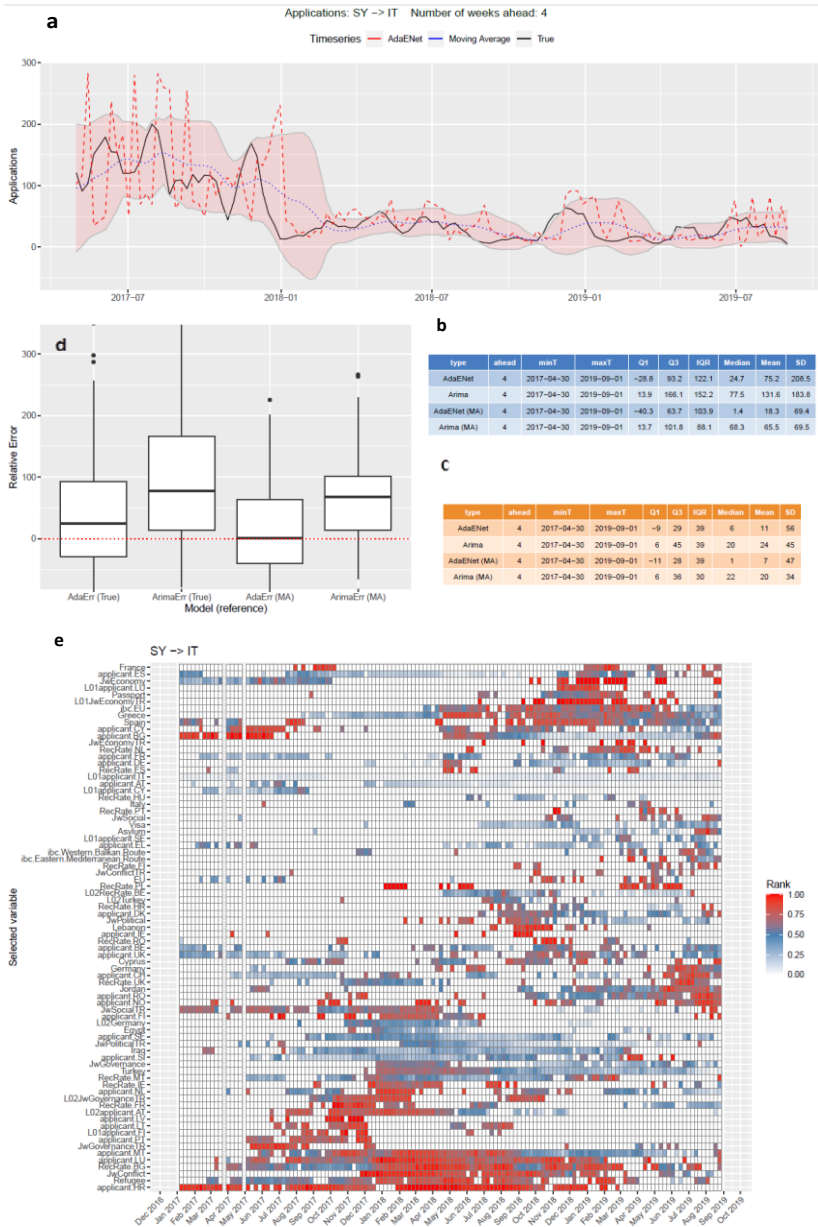

*Extended Data Fig. 2. Back-testing performance of the system for forecasted applications by Syrians in Italy.*

*a. The black line shows the actual number of applications lodged by Syrian nationals in Italy. The dotted blue line is the moving average of the process. The red dashed line shows the AdaENet 4-week ahead forecast at each time point. The pink shaded area represents a ± 2-standard errors confidence bands around the moving average.*

*b-c. Summary statistics for the relative error (b) and for the absolute error (c). Arima is a benchmark model which is only based on the auto-correlation of the application timeseries.*

*e. The factors that impact the forecasting model in the period considered. The model adapts over time, some effects are persistent in the first period, then others become more important. The scale colour only represents the relative importance of the variable. The coloured variables are all included in the model, the others have been dropped by AdaENet.*

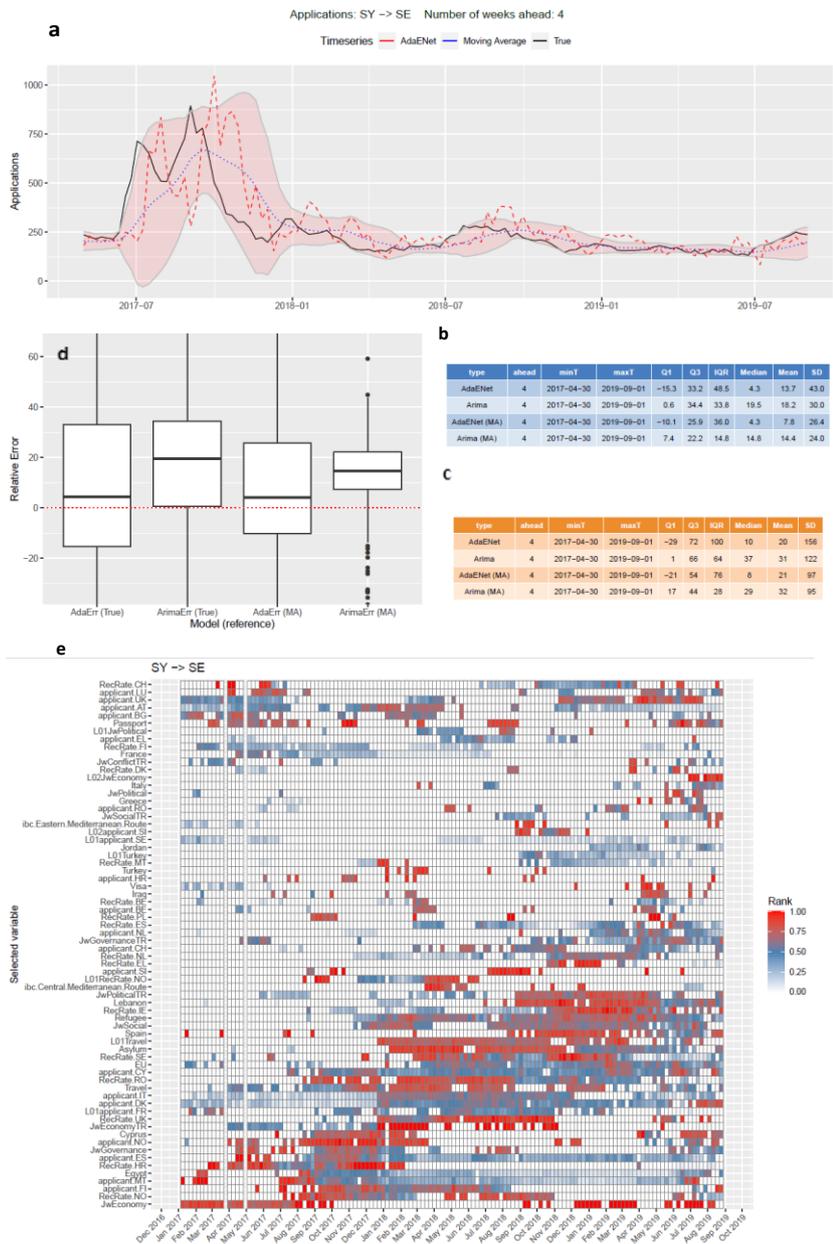

***Extended Data Fig. 3. Back-testing performance of the system for forecasted applications by Syrians in Sweden.***
***a.** The black line shows the actual number of applications lodged by Syrian nationals in Sweden. The dotted blue line is the moving average of the process. The red dashed line shows the AdaENet 4-week ahead forecast at each time point. The pink shaded area represents a ± 2-standard errors confidence bands around the moving average.*
***b-c**. Summary statistics for the relative error (**b**) and for the absolute error (**c**). Arima is a benchmark model which is only based on the auto-correlation of the application timeseries*
***e.** The factors that impact the forecasting model in the period considered. The model adapts over time, some effects are persistent in the first period, then others become more important. The scale colour only represents the relative importance of the variable. The coloured variables are all included in the model, the others have been dropped by AdaENet.*

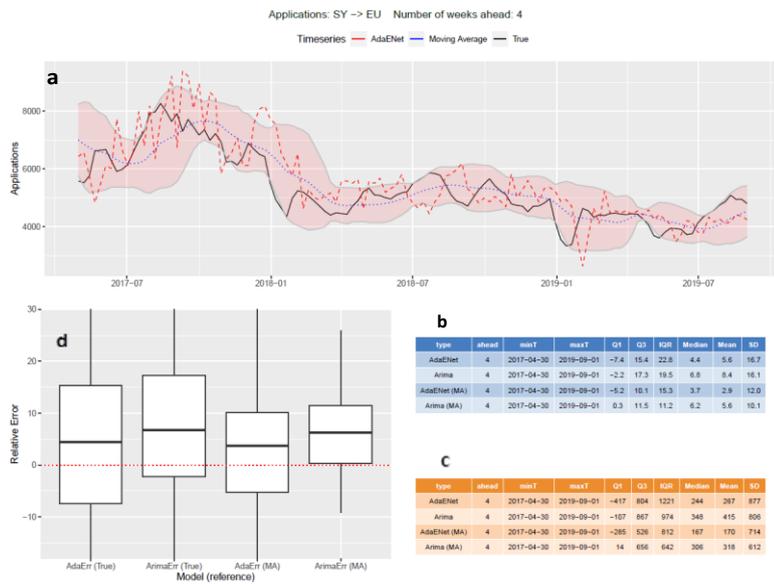

**Extended Data Fig. 4. Back-testing performance of the system for forecasted applications by Syrians in the EU+.**

*a.* The black line shows the actual number of applications lodged by Syrian nationals in the EU+. The dotted blue line is the moving average of the process. The red dashed line shows the AdaENet 4-week ahead forecast at each time point. The pink shaded area represents a ± 2-standard errors confidence bands around the moving average.

*b-c.* Summary statistics for the relative error (*b*) and for the absolute error (*c*). Arima is a benchmark model which is only based on the auto-correlation of the application timeseries

*e.* The factors that impact the forecasting model in the period considered. The model adapts over time, some effects are persistent in the first period, then others become more important. The scale colour only represents the relative importance of the variable. The coloured variables are all included in the model, the others have been dropped by AdaENet.

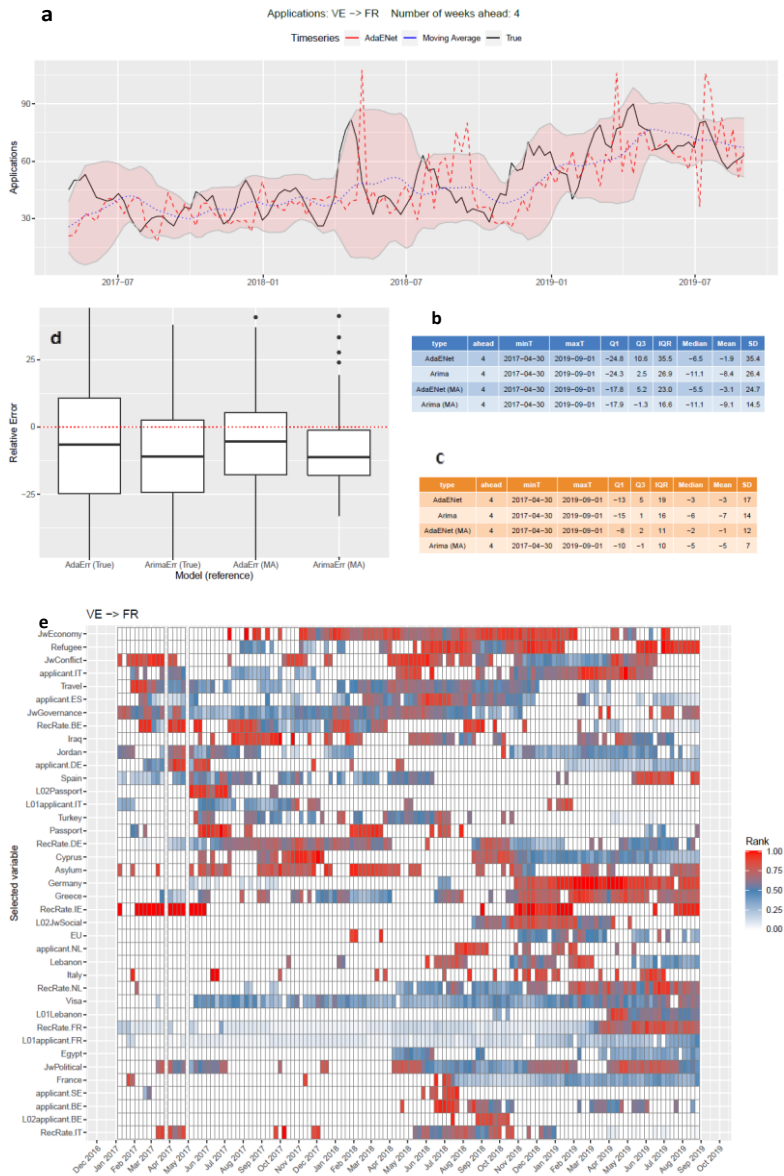

*Extended Data Fig. 5. Back-testing performance of the system for forecasted applications by Venezuelans in France.*

*a. The black line shows the actual number of applications lodged by Venezuelan nationals in France. The dotted blue line is the moving average of the process. The red dashed line shows the AdaENet 4-week ahead forecast at each time point. The pink shaded area represents a ± 2-standard errors confidence bands around the moving average.*

*b-c. Summary statistics for the relative error (b) and for the absolute error (c). Arima is a benchmark model which is only based on the auto-correlation of the application timeseries*

*e. The factors that impact the forecasting model in the period considered. The model adapts over time, some effects are persistent in the first period, then others become more important. The scale colour only represents the relative importance of the variable. The coloured variables are all included in the model, the others have been dropped by AdaENet.*

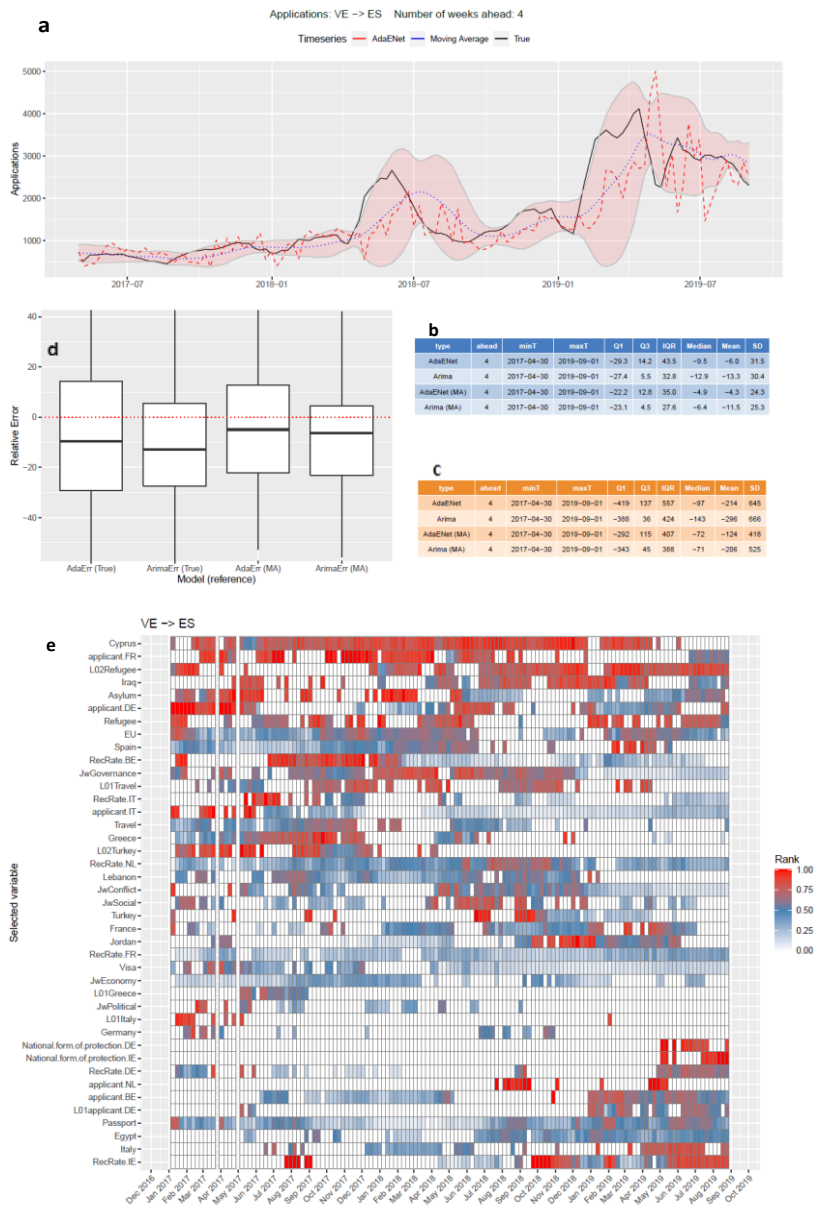

*Extended Data Fig. 6. Back-testing performance of the system for forecasted applications by Venezuelans in Spain.*

*a. The black line shows the actual number of applications lodged by Venezuelan nationals in Spain. The dotted blue line is the moving average of the process. The red dashed line shows the AdaENet 4-week ahead forecast at each time point. The pink shaded area represents a ± 2-standard errors confidence bands around the moving average.*

*b-c. Summary statistics for the relative error (b) and for the absolute error (c). Arima is a benchmark model which is only based on the auto-correlation of the application timeseries*

*e. The factors that impact the forecasting model in the period considered. The model adapts over time, some effects are persistent in the first period, then others become more important. The scale colour only represents the relative importance of the variable. The coloured variables are all included in the model, the others have been dropped by AdaENet.*

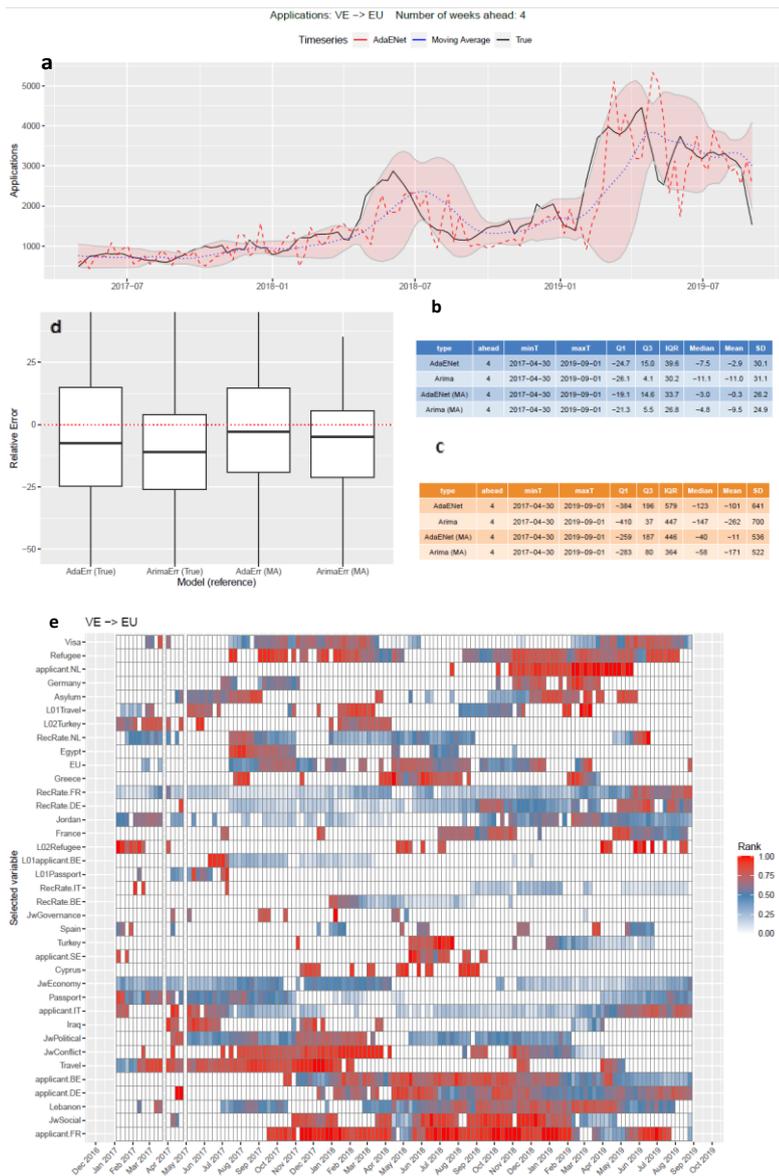

*Extended Data Fig. 7. Back-testing performance of the system for forecasted applications by Venezuelans in the EU+.*
*a. The black line shows the actual number of applications lodged by Venezuelan nationals in the EU+. The dotted blue line is the moving average of the process. The red dashed line shows the AdaENet 4-week ahead forecast at each time point. The pink shaded area represents a ± 2-standard errors confidence bands around the moving average.*
*b-c. Summary statistics for the relative error (b) and for the absolute error (c). Arima is a benchmark model which is only based on the auto-correlation of the application timeseries*
*e. The factors that impact the forecasting model in the period considered. The model adapts over time, some effects are persistent in the first period, then others become more important. The scale colour only represents the relative importance of the variable. The coloured variables are all included in the model, the others have been dropped by AdaENet.*

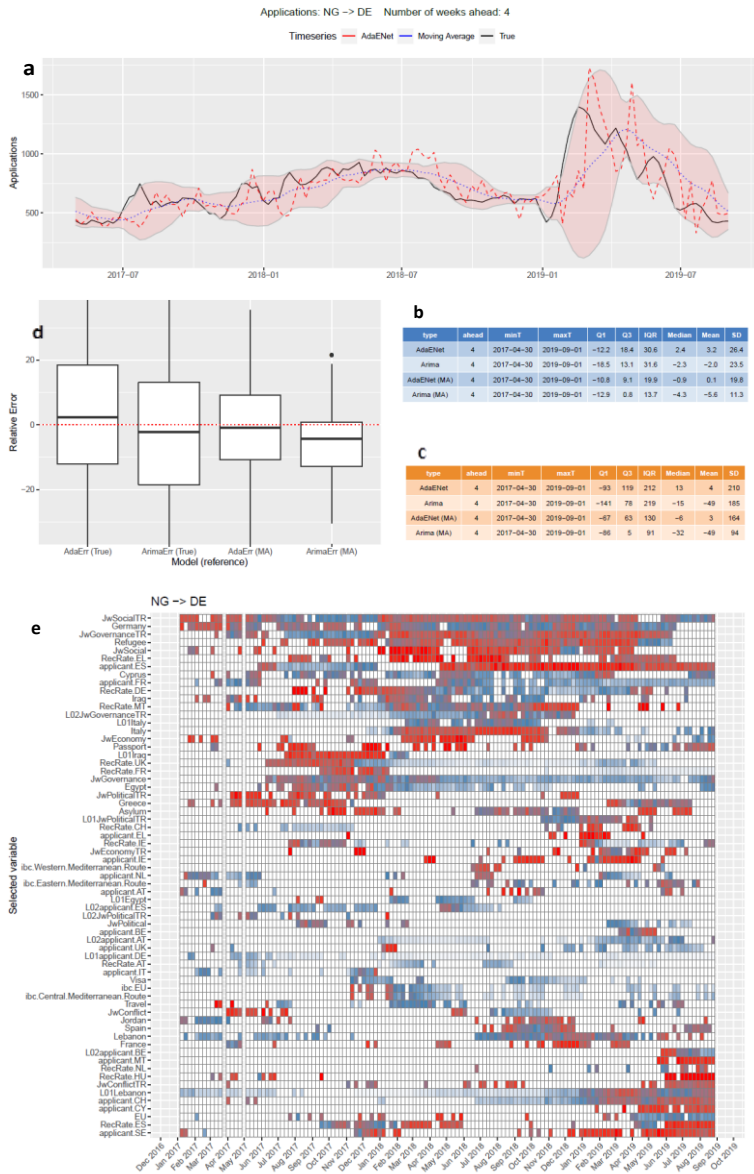

*Extended Data Fig. 8. Back-testing performance of the system for forecasted applications by Nigerians in Germany.*

***a.*** *The black line shows the actual number of applications lodged by Nigerian nationals in Germany. The dotted blue line is the moving average of the process. The red dashed line shows the AdaENet 4-week ahead forecast at each time point. The pink shaded area represents a ± 2-standard errors confidence bands around the moving average.*

***b-c.*** *Summary statistics for the relative error (**b**) and for the absolute error (**c**). Arima is a benchmark model which is only based on the auto-correlation of the application timeseries*

***e.*** *The factors that impact the forecasting model in the period considered. The model adapts over time, some effects are persistent in the first period, then others become more important. The scale colour only represents the relative importance of the variable. The coloured variables are all included in the model, the others have been dropped by AdaENet.*

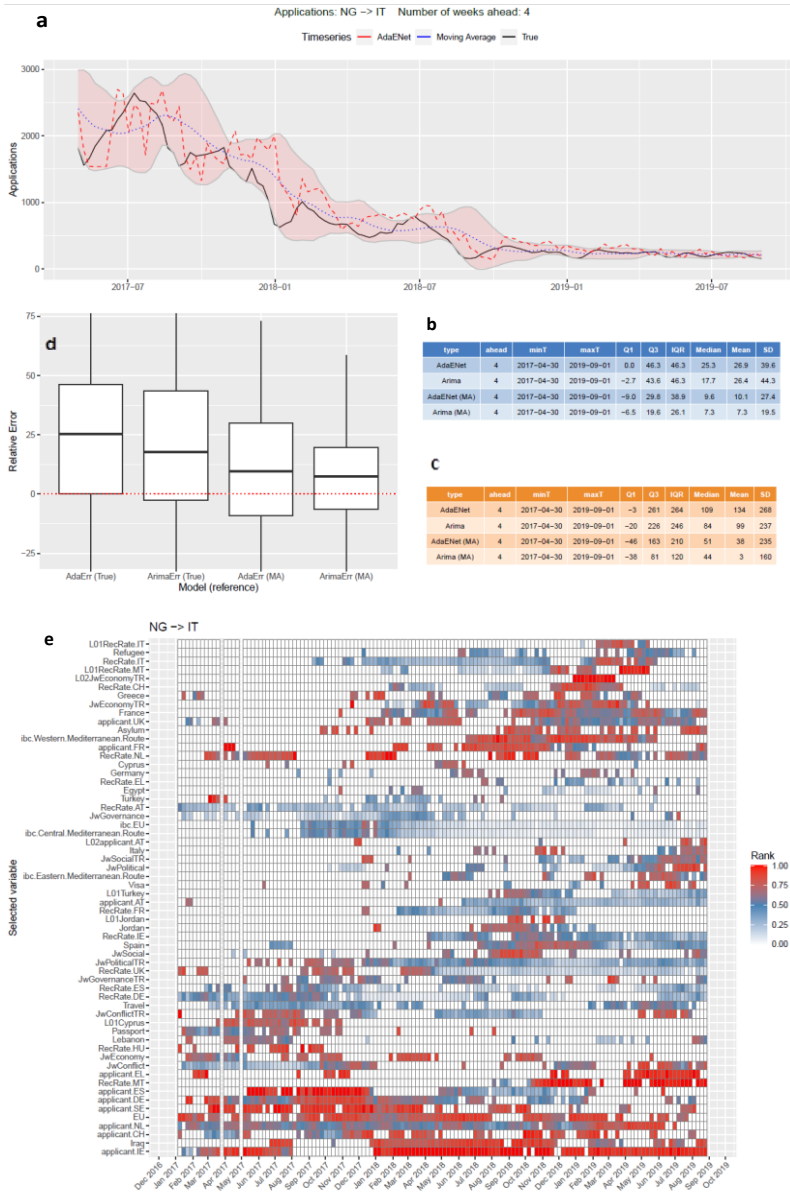

*Extended Data Fig. 9. Back-testing performance of the system for forecasted applications by Nigerians in Italy.*

*a. The black line shows the actual number of applications lodged by Nigerian nationals in Italy. The dotted blue line is the moving average of the process. The red dashed line shows the AdaENet 4-week ahead forecast at each time point. The pink shaded area represents a ± 2-standard errors confidence bands around the moving average.*

*b-c. Summary statistics for the relative error (b) and for the absolute error (c). Arima is a benchmark model which is only based on the auto-correlation of the application timeseries*

*e. The factors that impact the forecasting model in the period considered. The model adapts over time, some effects are persistent in the first period, then others become more important. The scale colour only represents the relative importance of the variable. The coloured variables are all included in the model, the others have been dropped by AdaENet.*

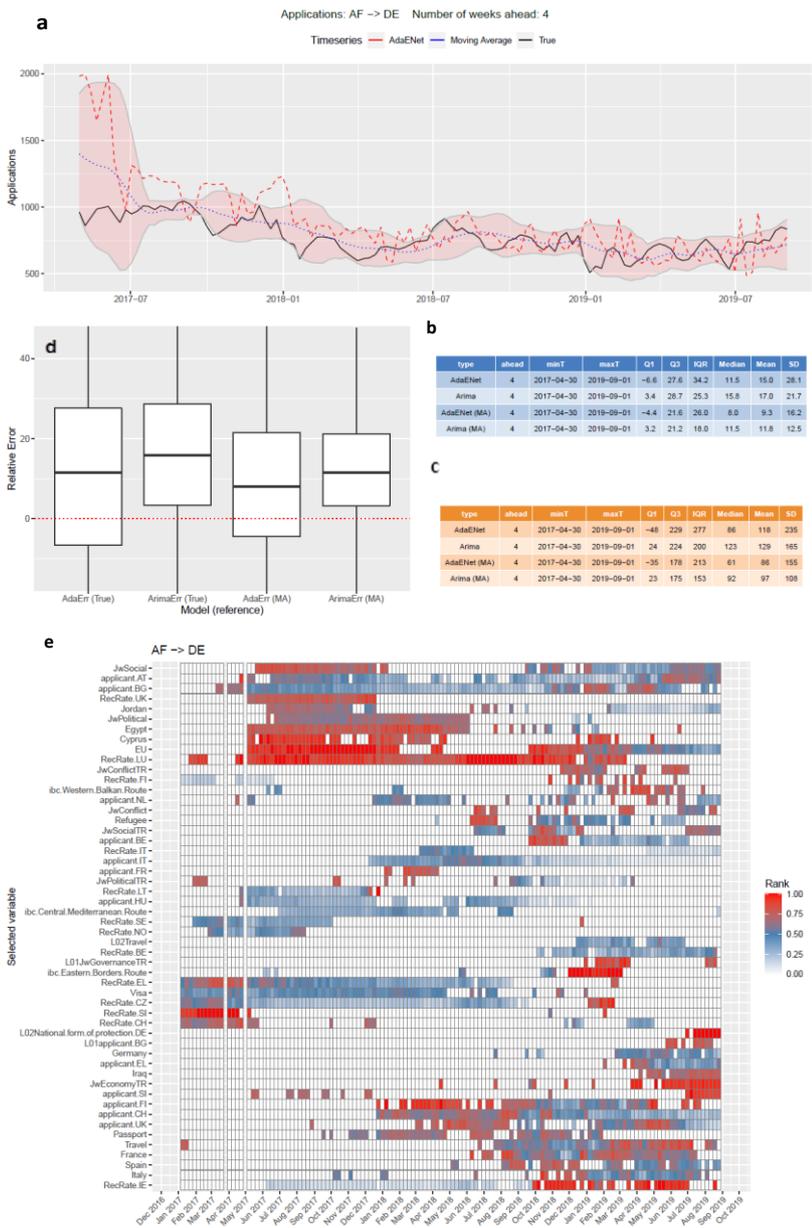

**Extended Data Fig. 10. Back-testing performance of the system for forecasted applications by Afghans in Germany.**
*a. The black line shows the actual number of applications lodged by Afghan nationals in Germany. The dotted blue line is the moving average of the process. The red dashed line shows the AdaENet 4-week ahead forecast at each time point. The pink shaded area represents a ± 2-standard errors confidence bands around the moving average.*
*b-c. Summary statistics for the relative error (b) and for the absolute error (c). Arima is a benchmark model which is only based on the auto-correlation of the application timeseries*
*e. The factors that impact the forecasting model in the period considered. The model adapts over time, some effects are persistent in the first period, then others become more important. The scale colour only represents the relative importance of the variable. The coloured variables are all included in the model, the others have been dropped by AdaENet.*

# Supplementary information

## Back-testing. Forecasting performance on selected country-of-origin-to-country-of-destination flows

We tested the performance of the system by simulating forecasts over a period of 118 weeks from 30 April 2017 to 1 September 2019, with the training period starting from April 2016 to allow at least a 50-week training set. The system starts by analysing data from April 2016 to April 2017, and then moves onward by one week at every step. The procedure replicates a hypothetical real forecast, that is, only information that would be available at each point in time is used; at each iteration, early warning analyses are ran to generate lagged variables that can be retained by the system in the forecasting step.

Note that the period selected for the back-test is a hard test for the forecasting system. In March 2016, the EU and Turkey agreed on stopping irregular migration on the Greek-Turkish border.[17] What the EU calls the 'Eastern Mediterranean route' was one of the main access points for mixed migration flows to Europe. Between April 2015 and March 2016, detections of irregular border-crossing were 85433 on the monthly average. In March 2016, after the EU-Turkey statement, there were 27343, and in April 2016 they went down to just 4324. Between April 2016 and March 2017, there were on the average 2864 detections of irregular border-crossing on the Eastern Mediterranean route every month.[18]

Because of the lag between detections of irregular border-crossing and the lodging of asylum applications, also due to the large backlog and delays that European asylum authorities were facing at that time, the level of applications went down a few months later. But it did so markedly: on the average, between October 2015 and September 2015, 123308 applications for asylum were lodged every month in EU countries; almost halved to an average of 63663 between October 2016 and September 2017. Thus, the period between 2016 and 2017 includes a radical change in the magnitude of mixed migration flows to the EU+ but not necessarily a change in push factors in countries of origin. To include that period in the back-testing makes the test rather conservative.

---

[17] See https://ec.europa.eu/commission/presscorner/detail/en/MEMO_16_963.
[18] Frontex data available at https://frontex.europa.eu/along-eu-borders/migratory-map/, last consulted on 1 July2020.

## Selection of countries and flows for back-testing

The early warning and forecasting system runs for all pairs of countries of origin (circa 200) to countries of destination (30 EU+ countries at the time of the analysis), which amounts to a total of circa 6000 dyads or country to country flows. Here we report results of back-testing on a selection of 70 dyads, generated by seven countries of origin (Afghanistan [AF], Eritrea [ER], Iraq [IQ], Nigeria [NG], Syria [SY], Turkey [TR] and Venezuela [VE]) and nine countries of destination (Austria [AT], Belgium [BE], Germany [DE], Greece [EL], Spain [ES], France [FR], Italy [IT], The Netherlands [NL] and Sweden [SE]); plus the EU+ as a whole.

The cases selected represent a suitably large diversity on the variables analysed. [19]

*Countries of origin*
- Global region and distance from countries of destination:
  - Countries in the Middle East (SY and TR), Asia (AF and PK), Africa (NG and ER) and America (VE)
- Internal situation during the period analysed:
  - Ongoing civil war (SY)
  - High (AF, IQ) to medium (PK, TR) fragility/instability
  - Authoritarian regime (ER)
  - Economic crises with relatively more limited incidence of conflicts events (NG, VE)
- Average rates of recognition of international protection in Europe[20]
  - Higher: ER (42.4% positive decisions; 44.7 including national forms of protection); SY (88.6% positive decisions; 89.7% including national forms of protection)
  - Medium: AF (35.8% positive decisions; 47.3% including national forms of protection); IQ (44.1% positive decisions; 47.6% including national forms of protection); VE (10.7% positive decisions; 77.9% including national forms of protection)[21]; TR (42.4% positive decisions; 44.7% including national forms of protection)

---

- - - o Lower: NG (9.5%; 20% including national forms of protection).
- Internet penetration[22] (Internet users: percentage, position in world rank)
    - o Medium-high (TR 65%, 92; VE 64%, 94)
    - o Medium-low (IQ 49%, 127)
    - o Low or very low (SY 34%, 153; NG 27%, 167; AF 199, 11%)
    - o World's lowest (ER 1%, 215)
- Outflows of asylum applicants to EU over the period analysed
    - o Consistently high levels (among top-3 to top-5 countries of origin): SY, IQ, AF
    - o Shifting from low-ranking to top-3: VE
    - o Occasionally high throughout the period: NG
    - o Relevant change: TR (never a top-country of origin but significantly high for some time)
    - o Consistently lower than the other countries of origin included: ER

Moreover, Turkey is contiguous with the EU i.e. Greece. In addition to being a relevant country of origin for part of the period, Turkey has been a major transit country for migration to Europe for most of the period used for training the model, as well as for the second part of the period analysed. It is currently the country hosting the most refugees in the world.

*Countries of destination*
- Geographic position:
    - o On the external border of the EU (EL/Eastern Mediterranean, ES/Western Mediterranean, IT/Central Mediterranean)
    - o Away from the external border, by increasing distance (AT, FR, DE, BE, NL, SE)
    - o Southern Europe (EL, ES, FR, IT), Central Europe (AT, DE), Northern Europe (BE, NL), Scandinavia (SE)
- Institutions and policy regimes for dealing with asylum-related migration
    - o Established (DE, FR, BE, NL, SE)
    - o Recent (EL, ES, IT)
- Asylum inflows over the period analysed
    - o Consistently in the top-five countries of destination: DE, EL, FR, IT
    - o From relatively low-ranking to top-3 country of destination: ES
    - o High levels relative to size (per-capita applications level): SE, BE, NL, AT (for some time)

*Flows*
- Level, variation, volatility, and patterns of asylum inflows:
    - High level throughout the whole period (AF, IQ, PK, SY)
    - o More (AF, IQ, PK) or relatively less (SY) variation
    - o Higher- (PK) or lower- (AF, IQ, SY) frequency change
    - o High in the first (ER, NG) or second (TR, VE) part of the observed period

---

general purpose here, however, is only to show that there is high variation in the recognition rates of applications
lodged by nationals of the set of countries included in the analysis.
[22] https://en.wikipedia.org/wiki/List_of_countries_by_number_of_Internet_users.

## Expectations

The high diversity represented in the sample of cases selected for back-testing provides a useful combination for assessing the adaptive capacity of the model. For example, events might be expected to be useful predictors of applications by nationals of such a country as Syria that has been in a civil war for years; or of Afghanistan, although to a lesser extent. Events should also be relevant to predicting asylum flows from countries that were unstable although not plagued by civil wars, like Pakistan; as well as from Nigeria and Venezuela afflicted by severe economic crises. Key transit countries can provide powerful barriers against migration to Europe, and in the back-testing we have included among covariates events in Turkey to account for that.

Internet searches, in turn, may be relevant to predicting migration from a range of countries (Bohme et al 2020), but less so from countries where internet penetration is limited. For some extreme cases like Eritrea, which is a stable dictatorship where the population has barely any access to the internet, events and internet searches are expected to be poor predictors. Our model, however, also includes variables related to EU border crossing as well as policy processes and outputs in countries of destination, such as applications and recognition rates in EU+ countries. These should be useful to support predictions when events or internet searches are not good predictors.

In general, the complexity of migration systems means that we cannot have strong expectations about the predictors of individual asylum flows. Even where we can conjecture about the relevance of individual factors to a given country to country flow, we still expect the system to change over time – and predictors to change accordingly. Therefore, we can only have some very general expectations about to the functioning and performance of the early warning and forecasting system, that we can group as follows:

1) Expectations related to the predictive capacity of the system:
- E1. The system should be able to predict asylum applications under different contexts
- E2. The forecasting error should normally stay within an acceptable confidence interval, i.e. +-2 standard errors from the real trend
- E3. Forecasts of relatively more stable, more patterned, and less volatile trends should have smaller errors than forecasts of more unstable, less patterned and more volatile trends
- E4. The Ada-ENet adaptive model should outperform a benchmark ARIMA model based on extrapolation from the trend

2) Expectations related to the capacity of the system to adapt to changing patterns:
- E5. Even in cases of unstable or volatile flows, the system should manage to quickly adapt to the trend.
- E6. In order to adapt to the functioning of individual migration systems, the system will select certain particular predictors and discard others. The selected predictors will vary across flows, as well as over time within the same flows to adapt to their dynamics.

## Results

Table A1. shows summary statistics describing the performance of the system over the entire back-testing period and across the selected country of origin to country of destination dyads and the related asylum flows. For each dyad, we show the relative and absolute average forecasting error. A1.1 includes results for the Ada-ENet model, while A1.2 includes results for the benchmark ARIMA model. To make sense of the results, we use a simple traffic light system, with colours indicating the forecasting performance ranging from very good (light green), to good (dark green), poor (orange) and very poor (red) performance. Moreover, a blue shading is added to those instances in which ARIMA outperforms Ada-ENet.

The ranges are based on a combination of relative and absolute average errors. We report both average and absolute errors because some high relative errors may still be very acceptable if absolute errors are small. For example, the 'very good performance' in the table includes those forecasts that have an average relative error smaller than 10% or an average absolute error smaller than 10. Because the forecasts are over a four-week horizon, an absolute error of 10 translates into just 2.5 asylum applications per week, on average. The double-thresholds is especially relevant to those flows that have a small magnitude, which means that forecasts can have a high relative but small absolute error. For similar reasons, flows of very small magnitude simply confound the assessment of the forecasting performance. Therefore, while the table shows all results for the sample of flows analysed, very small flows are marked grey to signal the limited or null significance to the testing exercise.

In addition to country to country flows, table A1 also includes flows from single countries of origin to the EU. We cannot use the same ranges designed for country-level flows to assess the performance of the system in forecasting EU-level asylum applications, therefore for the EU level we focus on relative errors only. All EU-level forecasts are very good except for the forecast of applications from Eritreans citizens, that is classified as good (but it only misses the 'very good' threshold by 0.3 points). Absolute errors for EU-level flows range between 63 and 267 (although the upper level is indeed an outlier, as the second-higher error is 149). When we divide those absolute errors by 30 EU+ countries, they become very low. Even if we divide absolute EU-level errors by 15 to account for the concentrated distribution of asylum applications across EU countries, absolute errors are still small. In any case, this is a theoretical exercise because errors related to single countries are computed separately.

When we exclude EU-level forecasts and the three cases where the numbers were not significant enough to produce any forecast, we are left with 59 country of origin to country of destination dyads. Of those, 16 still have very small magnitude and are marked grey. Therefore, we can comment on 43 dyads. Out of them, our Ada-ENet model performs very well in 30 cases, well in four cases, poorly in three cases, and very poorly in six cases. In other words, within our (highly diverse) sample, seven out of ten Ada-ENet forecasts are very good, and eight out of ten are very good or good; two out of ten are poor or very poor.

In comparison, the ARIMA model performs considerably worse (although perhaps still acceptably). Out of 43 dyads, the ARIMA produces 11 very good forecasts and 12 good forecasts. Poor and very poor forecasts are eight each. In all but four cases out of 43 (or in 90% of the observed cases) Ada-ENet outperforms the ARIMA model. Of the four dyads in which ARIMA performs better, one (Eritreans to Italy) is very poor with both models, and one (Syrians to The Netherlands) is very good with both models; therefore, we can consider the improvement of ARIMA on Ada-ENet as non-significant. Significant improvements can be seen with the forecasts of Syrians to Spain and Iraqis to Greece – respectively, from very poor to poor and from poor to good.

## Ada-ENet

| | Countries of destination | | | | | | | | | | | | | | | | | | | |
|---|---|---|---|---|---|---|---|---|---|---|---|---|---|---|---|---|---|---|---|---|
| | Relative | Absolute | Relative | Absolute | Relative | Absolute | Relative | Absolute | Relative | Absolute | Relative | Absolute | Relative | Absolute | Relative | Absolute | Relative | Absolute | Relative | Absolute |
| | AT | AT | BE | BE | DE | DE | EL | EL | ES | ES | FR | FR | IT | IT | NL | NL | SE | SE | EU | EU |
| SY | 14,1 | 38 | 3,6 | -7 | 7 | 113 | 13,2 | 31 | 47,5 | 60 | 18,8 | 45 | 75,2 | 11 | -2,7 | -12 | 13,7 | 20 | 5,6 | 267 |
| IQ | 20 | 9 | 6,2 | 7 | 1,5 | -6 | 11,3 | 28 | 17,3 | -1 | 9,5 | 3 | 15,6 | 8 | 8,5 | 3 | 13,2 | 13 | 5,9 | 149 |
| AF | 8,5 | 8 | 2,1 | -1 | 15 | 118 | 8,9 | 35 | | | -2,4 | -28 | 15,8 | 6 | 11,6 | 2 | 12,9 | 7 | 4,4 | 107 |
| NG | 13,9 | 0 | | | 3,2 | 4 | 12,9 | 0 | 36,3 | 5 | | | | | | | | | 5,5 | 67 |
| VE | | | 28,3 | 3 | -1,4 | -3 | | | -6 | -216 | -1,9 | -3 | 1,5 | -1 | 7,5 | -1 | 42,8 | 3 | -2,9 | -101 |
| TR | 20,5 | | 14,6 | 1 | 4,4 | 24 | 8,4 | -1 | 44,1 | 5 | -2,3 | -11 | 13,6 | 1 | 1,1 | -1 | 5,6 | 0 | -2 | -63 |
| ER | 31,7 | 1 | 22,8 | 2 | 16,3 | 42 | 50,5 | 5 | 46,1 | 1 | 1,9 | -1 | 100,8 | 23 | 10,2 | 3 | 41,1 | 2 | 10,3 | 109 |
| | | | | | | | | | | | | | | | | | | | | |
| Mean | 13,8 | 17,66667 | 3,966667 | -0,33333 | 6,571429 | 41,71429 | 10,45 | 23,25 | 20,75 | -77 | 3,3 | 0,142857 | 39,3 | 30,16667 | 4,15 | -1,83333 | 16,75 | 7,333333 | 3,828571 | 76,42857 |
| SD | 5,156549 | 17,61628 | 2,074448 | 7,023769 | 6,723272 | 53,09336 | 2,227854 | 16,41899 | 37,83021 | 193,7473 | 8,020598 | 22,22932 | 39,41472 | 51,472 | 6,789035 | 6,047038 | 12,34127 | 7,788881 | 4,684625 | 125,5586 |
| Max | 18,8 | 38 | 6,2 | 7 | 16,3 | 118 | 13,2 | 35 | 47,5 | 60 | 18,8 | 45 | 100,8 | 134 | 11,6 | 12 | 41,1 | 20 | 10,3 | 267 |
| Min | 8,5 | 7 | 2,1 | 1 | 1,4 | 3 | 8,4 | -1 | 5 | 1 | 0,5 | 3 | 1,5 | 1 | 2,7 | 1 | 5,6 | 2 | 2,9 | 67 |

## Arima

| | Countries of destination | | | | | | | | | | | | | | | | | | | |
|---|---|---|---|---|---|---|---|---|---|---|---|---|---|---|---|---|---|---|---|---|
| | Relative | Absolute | BE | BE | Relative | Absolute | Relative | Absolute | Relative | Absolute | Relative | Absolute | IT | IT | Relative | Absolute | SE | SE | Relative | Absolute |
| | AT | AT | BE | BE | DE | DE | EL | EL | ES | ES | FR | FR | IT | IT | NL | NL | SE | SE | EU | EU |
| SY | 14,3 | 37 | 6,4 | 0 | 15,4 | 301 | 18,5 | 80 | 29,2 | -11 | 24,6 | 58 | 131,6 | 24 | -1,9 | -10 | 18,2 | 31 | 8,4 | 415 |
| IQ | 29,3 | 12 | 1,3 | 1 | 10,9 | 98 | 5,5 | -12 | 40,2 | 0 | 10,6 | 2 | 15,4 | 6 | -7 | 15 | 14 | | 7,6 | 199 |
| AF | 16,3 | 27 | -2,9 | -12 | 17 | 129 | -5 | -77 | | | -6,3 | -70 | 16,8 | 6 | 14 | 4 | 17,3 | 13 | 4,8 | 115 |
| NG | 22,5 | 11 | -6,1 | -2 | | | 9,6 | 0 | 20,2 | -1 | -7,1 | -24 | 26,4 | 99 | -27,5 | -51 | 4,5 | 0 | 6,3 | 81 |
| VE | | | -2,7 | -5 | -8,3 | -6 | | | -13,1 | -296 | -8,4 | -7 | -8,9 | -10 | -12 | -4 | 2,5 | -1 | -11 | -262 |
| TR | 27,1 | 3 | 36,1 | 7 | -0,9 | -33 | 43,2 | 3 | 23,1 | 0 | -8,3 | -24 | 4 | -2 | -15,7 | -21 | 23,1 | 6 | -9,5 | -172 |
| ER | 29,4 | 1 | 8,3 | -10 | 23,3 | 5 | 43,2 | 3 | 33,7 | 1 | 1,6 | -3 | 189,6 | 56 | 21,9 | 11 | 52,4 | 4 | 8,1 | 63 |
| | | | | | | | | | | | | | | | | | | | | |
| Mean | 17,7 | 25 | 1,6 | -3,66667 | 7,871429 | 70,14286 | 4,525 | -8,75 | 7,95 | -153,5 | 0,957143 | -9,71429 | 61,81667 | 30,16667 | -1,2 | -12,3333 | 21,75 | 11,33333 | 2,1 | 62,71429 |
| SD | 4,275512 | 13,11488 | 4,657252 | 7,23178 | 11,66958 | 121,4721 | 10,26982 | 65,42871 | 30,05204 | 201,5254 | 12,53806 | 38,33499 | 79,5418 | 40,55819 | 18,31568 | 21,97878 | 16,22822 | 11,02119 | 8,534635 | 225,9341 |
| Max | 22,5 | 37 | 6,4 | 12 | 23,2 | 301 | 18,5 | 80 | 29,2 | 296 | 24,6 | 70 | 189,6 | 99 | 27,5 | 51 | 52,4 | 31 | 11 | 415 |
| Min | 14,3 | 11 | 1,3 | 0 | 2 | 6 | 5 | 0 | 13,3 | 0 | 6,3 | 2 | 8,9 | 6 | 2 | 4 | 4,5 | 0 | 4,8 | 81 |

## Traffic light system legend

| | | |
|---|---|---|
| Very good | | Rel error < 10% and/or abs error < 10 |
| Good | | Rel error < 10 and abs error > 20; OR rel err > 10 and abs < 20 |
| Poor | | Rel error > 10% and Abs error > 20 and < 40 |
| Very poor | | Rel error > 10% and Abs error > 40 |
| Arima better | | Arima better than Ada-ENet |

*Supplementary Information Table 1. Back-tested forecasting performance on selected representative flows. Summary statistics and comparison between Ada-ENet model and benchmark ARIMA model.*

*Performance of the system over the entire back-testing period and across the selected country of origin to country of destination dyads and the related asylum flows. For each dyad, the relative and absolute average forecasting errors are shown. A simple traffic light system is used for quick inspection of general performance results.*

*a. Ada-ENet. black line shows the actual number of applications lodged by Afghan nationals in Germany. b Arima xxx. c. Traffic-light system legend. Colours indicate general forecasting performance, ranging from very good (light green), to good (dark green), poor (orange) and very poor (red) performance. An additional blue shading denotes those instances in which ARIMA outperforms Ada-ENet.*

# Event categories. Selection and transformation

The following table includes all the event categories within the GDELT-CAMEO event codebook (See GDELT codebook http://data.gdeltproject.org/documentation/GDELT-Event_Codebook-V2.0.pdf and CAMEO scheme http://data.gdeltproject.org/documentation/CAMEO.Manual.1.1b3.pdf), as well as details on the construction of the event indices for the purpose of this study:

a) The events selected as potential drivers of migration (*GDELT code and topic*);

b) The *sign* associated to single events, denoting a potential to generate (+) or constrain (-) international displacements;

c) The grouping factor (*category label*) used to aggregate selected events into the five macro-categories defined for the purposes of this study: conflict, economic events, social unrest, governance-related events, political events;

d) The *strength* of the potential displacement-generation effect associated to single events.

| GDELT code and topic | Included | Sign | Category label | Strength |
|---|---|---|---|---|
| 110: Disapprove, not specified below | No | | | |
| 111: Criticize or denounce | No | | | |
| 112: Accuse, not specified below | No | | | |
| 0232: Appeal for military aid | Yes | + | Conflict | 1 |
| 0234: Appeal for military protection or peacekeeping | Yes | + | Conflict | 1 |
| 0254: Appeal for easing of economic sanctions, boycott, or embargo | Yes | + | Economic | 1 |
| 0255: Appeal for target to allow international involvement (non-mediation) | Yes | + | Conflict | 1 |
| 025: Social: Appeal for de-escalation of military engagement | Yes | + | Conflict | 1 |
| 02 Social: Appeal to others to meet or negotiate | Yes | + | Conflict | 1 |
| 027: Appeal to others to settle dispute | Yes | + | Conflict | 1 |
| 028: Appeal to engage in or accept mediation | Yes | + | Conflict | 1 |
| 0354: Express intent to ease economic sanctions, boycott, or embargo | Yes | - | Economic | 1 |
| 0355: Express intent to allow international involvement (non-mediation) | Yes | - | Conflict | 1 |

| GDELT code and topic | Included | Sign | Category label | Strength |
|---|---|---|---|---|
| 035Social: Express intent to de-escalate military engagement | Yes | - | Conflict | 1 |
| 03Social: Express intent to meet or negotiate | Yes | - | Conflict | 1 |
| 037: Express intent to settle dispute | Yes | - | Conflict | 1 |
| 038: Express intent to accept mediation | Yes | - | Conflict | 1 |
| 081: Ease administrative sanctions, not specified below | Yes | - | Conflict | 1 |
| 0871: Declare truce, ceasefire | Yes | - | Conflict | 1 |
| 0872: Ease military blockade | Yes | - | Conflict | 1 |
| 0873: Demobilize armed forces | Yes | - | Conflict | 2 |
| 0874: Retreat or surrender militarily | Yes | - | Conflict | 3 |
| 093: Investigate military action | Yes | + | Conflict | 2 |
| 094: Investigate war crimes | Yes | + | Conflict | 2 |
| 1012: Demand military cooperation | Yes | + | Conflict | 1 |
| 1014: Demand intelligence cooperation | Yes | + | Conflict | 1 |
| 1032: Demand military aid | Yes | + | Conflict | 1 |
| 1034: Demand military protection or peacekeeping | Yes | + | Conflict | 1 |
| 1054: Demand easing of economic sanctions, boycott, or embargo | Yes | + | Economic | 2 |
| 1055: Demand that target allows international involvement (non-mediation) | Yes | + | Conflict | 1 |
| 105Social: Demand de-escalation of military engagement | Yes | + | Conflict | 1 |
| 10Social: Demand meeting, negotiation | Yes | + | Conflict | 1 |
| 107: Demand settling of dispute | Yes | + | Conflict | 1 |
| 108: Demand mediation | Yes | + | Conflict | 1 |
| 1123: Accuse of aggression | Yes | + | Conflict | 1 |
| 1124: Accuse of war crimes | Yes | + | Conflict | 1 |
| 1125: Accuse of espionage, treason | Yes | + | Conflict | 1 |
| 1244: Refuse to ease economic sanctions, boycott, or embargo | Yes | + | Economic | 2 |
| 1245: Refuse to allow international involvement (non mediation) | Yes | + | Conflict | 2 |
| 124Social: Refuse to de-escalate military engagement | Yes | + | Conflict | 2 |
| 125: Reject proposal to meet, discuss, or negotiate | Yes | + | Conflict | 2 |
| 12Social: Reject mediation | Yes | + | Conflict | 2 |
| 127: Reject plan, agreement to settle dispute | Yes | + | Conflict | 2 |
| 138: Threaten with military force, not specified below | Yes | + | Conflict | 1 |
| 1382: Threaten occupation | Yes | + | Conflict | 2 |
| 1383: Threaten unconventional violence | Yes | + | Conflict | 2 |
| 1384: Threaten conventional attack | Yes | + | Conflict | 2 |
| 1385: Threaten attack with WMD | Yes | + | Conflict | 2 |
| 139: Give ultimatum | Yes | + | Conflict | 2 |
| 150: Demonstrate military or police power, not specified below | Yes | + | Conflict | 2 |
| 152: Increase military alert status | Yes | + | Conflict | 1 |

| GDELT code and topic | Included | Sign | Category label | Strength |
|---|---|---|---|---|
| 154: Mobilize or increase armed forces | Yes | + | Conflict | 2 |
| 155: Mobilize or increase cyber-forces | Yes | + | Conflict | 2 |
| 1Social4: Halt negotiations | Yes | + | Conflict | 2 |
| 1Social5: Halt mediation | Yes | + | Conflict | 1 |
| 1SocialSocial: Expel or withdraw, not specified below | Yes | + | Conflict | 2 |
| 1SocialSocial1: Expel or withdraw peacekeepers | Yes | + | Conflict | 2 |
| 1SocialSocial2: Expel or withdraw inspectors, observers | Yes | + | Conflict | 2 |
| 1SocialSocial3: Expel or withdraw aid agencies | Yes | + | Conflict | 2 |
| 170: Coerce, not specified below | Yes | + | Conflict | 2 |
| 17Social: Attack cybernetically | Yes | + | Conflict | 1 |
| 180: Use unconventional violence, not specified below | Yes | + | Conflict | 3 |
| 190: Use conventional military force, not specified below | Yes | + | Conflict | 3 |
| 191: Impose blockade, restrict movement | Yes | + | Conflict | 2 |
| 192: Occupy territory | Yes | + | Conflict | 3 |
| 193: Fight with small arms and light weapons | Yes | + | Conflict | 3 |
| 194: Fight with artillery and tanks | Yes | + | Conflict | 3 |
| 195: Employ aerial weapons, not specified below | Yes | + | Conflict | 3 |
| 1951: Employ precision-guided aerial munitions | Yes | + | Conflict | 3 |
| 1952: Employ remotely piloted aerial munitions | Yes | + | Conflict | 3 |
| 19Social: Violate ceasefire | Yes | + | Conflict | 3 |
| 200: Use unconventional mass violence, not specified below | Yes | + | Conflict | 4 |
| 204: Use weapons of mass destruction, not specified below | Yes | + | Conflict | 4 |
| 2041: Use chemical, biological, or radiological weapons | Yes | + | Conflict | 4 |
| 2042: Detonate nuclear weapons | Yes | + | Conflict | 4 |
| 1033: Demand humanitarian aid | Yes | + | Conflict | 2 |
| 023: Appeal for aid, not specified below | Yes | + | Economic | 2 |
| 0231: Appeal for economic aid | Yes | + | Economic | 2 |
| 1011: Demand economic cooperation | Yes | + | Economic | 2 |
| 103: Demand material aid, not specified below | Yes | + | Economic | 2 |
| 1031: Demand economic aid | Yes | + | Economic | 2 |
| 024: Appeal for political reform, not specified below | Yes | + | Governance | 1 |
| 0241: Appeal for change in leadership | Yes | + | Governance | 1 |
| 0244: Appeal for change in institutions, regime | Yes | + | Governance | 1 |
| 034: Express intent to institute political reform, not specified below | Yes | - | Governance | 1 |
| 0341: Express intent to change leadership | Yes | - | Governance | 1 |
| 0342: Express intent to change policy | Yes | - | Governance | 1 |
| 0344: Express intent to change institutions, regime | Yes | - | Governance | 1 |

| GDELT code and topic | Included | Sign | Category label | Strength |
|---|---|---|---|---|
| 035: Express intent to yield, not specified below | Yes | - | Governance | 1 |
| 083: Accede to requests or demands for political reform, not specified below | Yes | - | Governance | 2 |
| 0831: Accede to demands for change in leadership | Yes | - | Governance | 2 |
| 0832: Accede to demands for change in policy | Yes | - | Governance | 2 |
| 0834: Accede to demands for change in institutions, regime | Yes | - | Governance | 2 |
| 091: Investigate crime, corruption | Yes | + | Governance | 1 |
| 104: Demand political reform, not specified below | Yes | + | Governance | 1 |
| 1041: Demand change in leadership | Yes | + | Governance | 1 |
| 1042: Demand policy change | Yes | + | Governance | 1 |
| 1044: Demand change in institutions, regime | Yes | - | Governance | 1 |
| 105: Demand that target yields, not specified below | Yes | + | Governance | 1 |
| 1121: Accuse of crime, corruption | Yes | + | Governance | 1 |
| 123: Reject request or demand for political reform, not specified below | Yes | + | Governance | 1 |
| 1231: Reject request for change in leadership | Yes | + | Governance | 1 |
| 1232: Reject request for policy change | Yes | + | Governance | 1 |
| 1234: Reject request for change in institutions, regime | Yes | + | Governance | 1 |
| 124: Refuse to yield, not specified below | Yes | + | Governance | 1 |
| 1241: Refuse to ease administrative sanctions | Yes | + | Governance | 1 |
| 128: Defy norms, law | Yes | + | Governance | 1 |
| 0243: Appeal for rights | Yes | + | Political | 1 |
| 0251: Appeal for easing of administrative sanctions | Yes | + | Political | 1 |
| 0253: Appeal for release of persons or property | Yes | + | Political | 1 |
| 0343: Express intent to provide rights | Yes | - | Political | 1 |
| 0351: Express intent to ease administrative sanctions | Yes | - | Political | 1 |
| 0353: Express intent to release persons or property | Yes | - | Political | 1 |
| 075: Grant asylum | Yes | - | Political | 2 |
| 0811: Ease restrictions on political freedoms | Yes | - | Political | 2 |

| GDELT code and topic | Included | Sign | Category label | Strength |
|---|---|---|---|---|
| 0812: Ease ban on political parties or politicians | Yes | - | Political | 2 |
| 0813: Ease curfew | Yes | - | Political | 2 |
| 0814: Ease state of emergency or martial law | Yes | - | Political | 2 |
| 0833: Accede to demands for rights | Yes | - | Political | 2 |
| 092: Investigate human rights abuses | Yes | + | Political | 1 |
| 1043: Demand rights | Yes | + | Political | 1 |
| 1051: Demand easing of administrative sanctions | Yes | + | Political | 1 |
| 1053: Demand release of persons or property | Yes | + | Political | 1 |
| 1122: Accuse of human rights abuses | Yes | + | Political | 1 |
| 1233: Reject request for rights | Yes | + | Political | 1 |
| 1243: Refuse to release persons or property | Yes | + | Political | 2 |
| 1322: Threaten to ban political parties or politicians | Yes | + | Political | 1 |
| 1323: Threaten to impose curfew | Yes | + | Political | 1 |
| 1324: Threaten to impose state of emergency or martial law | Yes | + | Political | 1 |
| 137: Threaten with repression | Yes | + | Political | 1 |
| 151: Increase police alert status | Yes | + | Political | 2 |
| 153: Mobilize or increase police power | Yes | + | Political | 2 |
| 1711: Confiscate property | Yes | + | Political | 3 |
| 172: Impose administrative sanctions, not specified below | Yes | + | Political | 3 |
| 1721: Impose restrictions on political freedoms | Yes | + | Political | 3 |
| 1722: Ban political parties or politicians | Yes | + | Political | 3 |
| 1723: Impose curfew | Yes | + | Political | 3 |
| 1724: Impose state of emergency or martial law | Yes | + | Political | 3 |
| 173: Arrest, detain, or charge with legal action | Yes | + | Political | 3 |
| 174: Expel or deport individuals | Yes | + | Political | 3 |
| 175: Use tactics of violent repression | Yes | + | Political | 3 |
| 181: Abduct, hijack, or take hostage | Yes | + | Political | 3 |
| 1822: Torture | Yes | + | Political | 3 |
| 1823: Kill by physical assault | Yes | + | Political | 3 |
| 201: Engage in mass expulsion | Yes | + | Conflict | 3 |
| 202: Engage in mass killings | Yes | + | Conflict | 3 |
| 203: Engage in ethnic cleansing | Yes | + | Conflict | 3 |
| 0233: Appeal for humanitarian aid | Yes | + | Conflict | 1 |
| 0252: Appeal for easing of political dissent | Yes | + | Social | 1 |
| 0352: Express intent to ease popular dissent | Yes | - | Social | 1 |
| 082: Ease political dissent | Yes | - | Social | 2 |
| 084: Return, release, not specified below | Yes | - | Political | 2 |
| 0841: Return, release person(s) | Yes | - | Political | 2 |
| 0842: Return, release property | Yes | - | Political | 2 |
| 1052: Demand easing of political dissent | Yes | + | Political | 1 |
| 113: Rally opposition against | Yes | + | Social | 2 |
| 1242: Refuse to ease popular dissent | Yes | + | Social | 2 |

| GDELT code and topic | Included | Sign | Category label | Strength |
|---|---|---|---|---|
| 133: Threaten with political dissent, protest | Yes | + | Social | 1 |
| 1381: Threaten blockade | Yes | + | Social | 1 |
| 140: Engage in political dissent, not specified below | Yes | + | Social | 1 |
| 141: Demonstrate or rally, not specified below | Yes | + | Social | 1 |
| 1411: Demonstrate for leadership change | Yes | + | Social | 2 |
| 1412: Demonstrate for policy change | Yes | + | Social | 1 |
| 1413: Demonstrate for rights | Yes | + | Social | 1 |
| 1414: Demonstrate for change in institutions, regime | Yes | + | Social | 2 |
| 142: Conduct hunger strike, not specified below | Yes | + | Social | 1 |
| 1421: Conduct hunger strike for leadership change | Yes | + | Social | 2 |
| 1422: Conduct hunger strike for policy change | Yes | + | Social | 1 |
| 1423: Conduct hunger strike for rights | Yes | + | Social | 1 |
| 1424: Conduct hunger strike for change in institutions, regime | Yes | + | Social | 2 |
| 143: Conduct strike or boycott, not specified below | Yes | + | Social | 1 |
| 1431: Conduct strike or boycott for leadership change | Yes | + | Social | 2 |
| 1432: Conduct strike or boycott for policy change | Yes | + | Social | 1 |
| 1433: Conduct strike or boycott for rights | Yes | + | Social | 1 |
| 1434: Conduct strike or boycott for change in institutions, regime | Yes | + | Social | 2 |
| 144: Obstruct passage, block, not specified below | Yes | + | Social | 1 |
| 1441: Obstruct passage to demand leadership change | Yes | + | Social | 2 |
| 1442: Obstruct passage to demand policy change | Yes | + | Social | 1 |
| 1443: Obstruct passage to demand rights | Yes | + | Social | 1 |
| 1444: Obstruct passage to demand change in institutions, regime | Yes | + | Social | 2 |
| 145: Protest violently, riot, not specified below | Yes | + | Social | 2 |
| 1451: Engage in violent protest for leadership change | Yes | + | Social | 3 |
| 1452: Engage in violent protest for policy change | Yes | + | Social | 3 |
| 1453: Engage in violent protest for rights | Yes | + | Social | 3 |
| 1454: Engage in violent protest for change in institutions, regime | Yes | + | Social | 3 |
| 171: Seize or damage property, not specified below | Yes | + | Social | 1 |
| 1712: Destroy property | Yes | + | Social | 1 |
| 182: Physically assault, not specified below | Yes | + | Social | 2 |
| 1821: Sexually assault | Yes | + | Social | 1 |
| 183: Conduct suicide, car, or other non-military bombing, not specified below | Yes | + | Social | 3 |
| 1831: Carry out suicide bombing | Yes | + | Social | 3 |
| 1832: Carry out vehicular bombing | Yes | + | Social | 3 |
| 1833: Carry out roadside bombing | Yes | + | Social | 3 |
| 1834: Carry out location bombing | Yes | + | Social | 1 |
| 184: Use as human shield | Yes | + | Social | 3 |
| 185: Attempt to assassinate | Yes | + | Social | 2 |

| GDELT code and topic | Included | Sign | Category label | Strength |
|---|---|---|---|---|
| 18Social: Assassinate | Yes | + | Social | 2 |
| 012: Make pessimistic comment | No | | | |
| 015: Acknowledge or claim responsibility | No | | | |
| 0312: Express intent to cooperate militarily | No | | | |
| 07: PROVIDE AID | Yes | - | Conflict | 1 |
| 070: Provide aid, not specified below | Yes | + | Conflict | 2 |
| 071: Provide economic aid | Yes | - | Economic | 2 |
| 072: Provide military aid | Yes | + | Conflict | 2 |
| 073: Provide humanitarian aid | Yes | - | Conflict | 2 |
| 074: Provide military protection or peacekeeping | Yes | - | Conflict | 2 |
| 085: Ease economic sanctions, boycott, embargo | Yes | - | Conflict | 2 |
| 08Social: Allow international involvement, not specified below | Yes | - | Conflict | 2 |
| 08Social1: Receive deployment of peacekeepers | Yes | - | Conflict | 2 |
| 08Social2: Receive inspectors | Yes | - | Conflict | 1 |
| 08Social3: Allow humanitarian access | Yes | - | Conflict | 2 |
| 12: REJECT | No | | | |
| 13: THREATEN | Yes | + | Conflict | 1 |
| 1Social: REDUCE RELATIONS | Yes | + | Conflict | 1 |
| 01: MAKE PUBLIC STATEMENT | No | | | |
| 010: Make statement, not specified below | No | | | |
| 011: Decline comment | No | | | |
| 013: Make optimistic comment | No | | | |
| 014: Consider policy option | No | | | |
| 01Social: Deny responsibility | No | | | |
| 017: Engage in symbolic act | No | | | |
| 018: Make empathetic comment | No | | | |
| 019: Express accord | No | | | |
| 02: APPEAL | No | | | |
| 020: Make an appeal or request, not specified below | No | | | |
| 021: Appeal for material cooperation, not specified below | No | | | |
| 0211: Appeal for economic cooperation | Yes | + | Economic | 1 |
| 0212: Appeal for military cooperation | Yes | + | Conflict | 1 |
| 0213: Appeal for judicial cooperation | No | | | |
| 0214: Appeal for intelligence | No | | | |
| 022: Appeal for diplomatic cooperation (such as policy support) | No | | | |
| 0242: Appeal for policy change | No | | | |
| 025: Appeal to yield, not specified below | No | | | |
| 03: EXPRESS INTENT TO COOPERATE | No | | | |
| 030: Express intent to cooperate, not specified below | No | | | |
| 031: Express intent to engage in material cooperation, not specified below | No | | | |

| GDELT code and topic | Included | Sign | Category label | Strength |
|---|---|---|---|---|
| 0311: Express intent to cooperate economically | Yes | - | Economic | 1 |
| 0313: Express intent to cooperate on judicial matters | No | | | |
| 0314: Express intent to cooperate on intelligence | No | | | |
| 032: Express intent to engage in diplomatic cooperation (such as policy support) | Yes | - | Conflict | 1 |
| 033: Express intent to provide material aid, not specified below | Yes | - | Conflict | 1 |
| 0331: Express intent to provide economic aid | Yes | - | Economic | 1 |
| 0332: Express intent to provide military aid | Yes | + | Conflict | 1 |
| 0333: Express intent to provide humanitarian aid | Yes | - | Conflict | 1 |
| 0334: Express intent to provide military protection or peacekeeping | Yes | - | Conflict | 1 |
| 039: Express intent to mediate | Yes | - | Conflict | 1 |
| 04: CONSULT | No | | | |
| 040: Consult, not specified below | No | | | |
| 041: Discuss by telephone | No | | | |
| 042: Make a visit | No | | | |
| 043: Host a visit | No | | | |
| 044: Meet at a "third" location | No | | | |
| 045: Mediate | No | | | |
| 04Social: Engage in negotiation | Yes | - | Conflict | 2 |
| 05: ENGAGE IN DIPLOMATIC COOPERATION | No | | | |
| 050: Engage in diplomatic cooperation, not specified below | Yes | - | Conflict | 2 |
| 051: Praise or endorse | No | | | |
| 052: Defend verbally | No | | | |
| 053: Rally support on behalf of | No | | | |
| 054: Grant diplomatic recognition | Yes | - | Conflict | 1 |
| 055: Apologize | No | | | |
| 05Social: Forgive | No | | | |
| 057: Sign formal agreement | No | | | |
| 0Social: ENGAGE IN MATERIAL COOPERATION | No | | | |
| 0Social0: Engage in material cooperation, not specified below | No | | | |
| 0Social1: Cooperate economically | Yes | - | Economic | 2 |
| 0Social2: Cooperate militarily | Yes | + | Conflict | 2 |
| 0Social3: Engage in judicial cooperation | No | | | |
| 0Social4: Share intelligence or information | No | | | |
| 08: YIELD | No | | | |
| 080: Yield, not specified below | No | | | |
| 087: De-escalate military engagement | Yes | - | Conflict | 3 |
| 09: INVESTIGATE | No | | | |
| 090: Investigate, not specified below | No | | | |
| 10: DEMAND | No | | | |
| 100: Demand, not specified below | No | | | |

| GDELT code and topic | Included | Sign | Category label | Strength |
|---|---|---|---|---|
| 101: Demand material cooperation, not specified below | No | | | |
| 1013: Demand judicial cooperation | No | | | |
| 102: Demand diplomatic cooperation (such as policy support) | No | | | |
| 11: DISAPPROVE | No | | | |
| 114: Complain officially | No | | | |
| 115: Bring lawsuit against | No | | | |
| 11Social: Find guilty or liable (legally) | No | | | |
| 120: Reject, not specified below | No | | | |
| 121: Reject material cooperation | No | | | |
| 1211: Reject economic cooperation | Yes | + | Economic | 1 |
| 1212: Reject military cooperation | Yes | + | Conflict | 1 |
| 122: Reject request or demand for material aid, not specified below | Yes | + | Conflict | 1 |
| 1221: Reject request for economic aid | Yes | + | Conflict | 1 |
| 1222: Reject request for military aid | Yes | + | Conflict | 1 |
| 1223: Reject request for humanitarian aid | Yes | + | Conflict | 1 |
| 1224: Reject request for military protection or peacekeeping | Yes | + | Conflict | 1 |
| 129: Veto | No | | | |
| 130: Threaten, not specified below | No | | | |
| 131: Threaten non-force, not specified below | No | | | |
| 1311: Threaten to reduce or stop aid | No | | | |
| 1312: Threaten with sanctions, boycott, embargo | No | | | |
| 1313: Threaten to reduce or break relations | No | | | |
| 132: Threaten with administrative sanctions, not specified below | No | | | |
| 1321: Threaten with restrictions on political freedoms | No | | | |
| 134: Threaten to halt negotiations | No | | | |
| 135: Threaten to halt mediation | No | | | |
| 13Social: Threaten to halt international involvement (non-mediation) | No | | | |
| 14: PROTEST | Yes | + | Social | 1 |
| 15: EXHIBIT FORCE POSTURE | Yes | + | Conflict | 1 |
| 1Social0: Reduce relations, not specified below | No | | | |
| 1Social1: Reduce or break diplomatic relations | No | | | |
| 1Social2: Reduce or stop material aid, not specified below | Yes | + | Conflict | 2 |
| 1Social21: Reduce or stop economic assistance | Yes | + | Economic | 2 |
| 1Social22: Reduce or stop military assistance | Yes | + | Conflict | 2 |
| 1Social23: Reduce or stop humanitarian assistance | Yes | + | Conflict | 2 |
| 1Social3: Impose embargo, boycott, or sanctions | Yes | + | Conflict | 2 |
| 17: COERCE | No | | | |
| 18: ASSAULT | No | | | |
| 19: FIGHT | No | | | |
| 20: USE UNCONVENTIONAL MASS VIOLENCE | No | | | |

**Table M1. Parameters for the Early Warning function.** Event categories within the GDELT-CAMEO event codebook, and details on the construction of the event indices for the purpose of this study:

- The events selected as potential drivers of migration (*GDELT code and topic*);
- The *sign* associated to single events, denoting a potential to generate (+) or constrain (-) international displacements;
- The grouping factor (*category label*) used to aggregate selected events into the five macro-categories defined for the purposes of this study: conflict, economic events, social unrest, governance-related events, political events;
- The *strength* of the potential displacement-generation effect associated to single events.